# The rise and fall of cosmical physics: notes for a history, ca. 1850-1920

Helge Kragh*

**Abstract**. In the period from about 1890 to 1915 an interdisciplinary and unifying research programme known as "cosmical physics" attracted much scientific and public attention. It typically included aspects of the earth sciences (such as magnetic storms and atmospheric electricity) combined with astronomical subjects (such as the solar corona and cometary tails), but there was no unanimity as to the precise meaning of cosmical physics, which collapsed after World War I. The essay covers the history of cosmical physics as it unfolded in particular in Germany, Austria, England and Scandinavia. Among the scientists who contributed to the development of cosmical physics were Wilhelm Förster from Germany, Wilhelm Trabert from Austria, Kristian Birkeland from Norway and Svante Arrhenius from Sweden. While cosmical physics did not usually involve a cosmological dimension, both Birkeland and Arrhenius constructed cosmologies based on their work on auroral and other phenomena.

In the years around 1900 there was much interest in *kosmische Physik*, or what in the English-speaking world was called *cosmical physics*, an ambitious attempt to create a synthetic research programme covering essential parts of the earth sciences and the astronomical sciences.[1] The central research problems of cosmical physics were often taken to be terrestrial phenomena in relation to the solar system, including geomagnetic storms, the aurora borealis, meteorites and atmospheric electricity, but there was no agreement as to what the term signified, more precisely.

"In recent time few branches of science have made such great progress as cosmical physics," wrote the Swedish chemist and physicist Svante Arrhenius in 1903. *The Times* repeated that cosmical physics "is a department of science which is … growing rapidly and healthily."[2] The sentiment was shared by many meteorologists, geophysicists and astronomers who saw a bright future for the new interdisciplinary field. But this was not what happened. While in the first decade of the new century cosmical physics was widely received with enthusiasm, after 1920 little more was heard of it. Cosmical physics was a promising research programme that failed to develop into a discipline and whose impact on science in the period about 1890 to 1915 was more rhetorical than real. Today this chapter in the history of science is long forgotten except by a few historians. Nonetheless, it is an interesting chapter that deserves to be known by both scientists and historians of science.

---

* Centre for Science Studies, Department of Physics and Astronomy, Aarhus University, Aarhus, Denmark. E-mail: helge.kragh@ivs.au.dk.

1 In the period under consideration the term "cosmical physics" was generally used rather than "cosmic physics." The first term is rarely used today and there is no entry for it in the *Oxford English Dictionary* (electronic version). On the other hand, scientists sometimes refer to "cosmic physics," but then in a meaning different from the one of the past.

2 Arrhenius 1903, p. iii. *The Times*, 27 December 1904.

The development of cosmical physics has sometimes been dealt with from more specific and local perspectives, especially in connection with Scandinavian science in the early years of the twentieth century.[3] Other authors have referred to the topic in disciplinary connections, such as the troubled relationship between meteorology and physics, but the development of cosmical physics has never been the subject of a comprehensive historical investigation across disciplines and nations. This essay is an attempt to provide as much material as possible for such an investigation.

Cosmical physics was a European and to some extent American adventure, but apparently it caught on only in a few countries, the most important of which were Germany, Austria, Switzerland, Sweden, Norway, England and the United States. For lack of knowledge of cosmical physics elsewhere, my study is limited to these countries. Of course, there were also scientists outside these countries who cultivated aspects of cosmical physics, but they did not, as far as I know, consider their work to be within a new disciplinary framework. To mention but one example, the Italian Jesuit scientist Angelo Secchi is best known as a pioneer of astrophysics, but in the 1860s and 1870s he also did important work in meteorology and geomagnetism, aiming in his research to correlate the physics of the heavens with that of the Earth. He may well be considered an early cosmical physicist, although the label might have been foreign to him. The same is the case with several of the scientists mentioned in this study who engaged in cosmical physics without using the term. Rather than adopting a chronological approach, I have structured my survey of cosmical physics nation-wise, starting with Germany and ending with the United States. While such a structure is not generally recommendable, in this case it is defensible because cosmical physics was perceived differently in the various nations. When Austrian, British and Swedish scientists spoke of cosmical physics, they did not speak of the same thing.

[3] See in particular Crawford 1992, Crawford 1996 and Holmberg 1999.

The content of the essay is the following: (1) Germany; (2) Switzerland; (3) Austria; (4) Scandinavia: Norway; (5) Scandinavia: Sweden; (6) Nobel contexts; (7) Arrhenius' cosmological speculations; (8) Great Britain; (9) The United States; (10) Cosmical physics after 1920; (11) Summary and discussion.

**1. Germany**

Alexander von Humboldt's encyclopaedic and hugely popular work *Kosmos*, published in five large volumes between 1845 and 1862 (the last one posthumously), served as an important source of inspiration for German writers and scientists in the cosmophysical tradition. Humboldt's guiding theme was a holistic view of the universe in which the seemingly chaotic phenomena of the terrestrial world were governed by the same universal laws that secured order and beauty in the heavens. The kind of science he advocated was empirical, quantitative, synthetic and thoroughly based on the use of instruments. Susan Faye Cannon defines it as "astronomy and the physics of the earth and the biology of the earth all viewed from a geographical standpoint, with the goal of discovering quantitative mathematical connections and interrelationships."[4]

Although Humboldt spoke of a comprehensive "terrestrial physics," in *Kosmos* he did not use the term "cosmical physics." In Germany, Otto Ule, a prolific writer of popular books and articles on science, was among those who followed in the footsteps of the great Humboldt. His *Die Natur* of 1851, written "in the spirit of Humboldt," covered an extremely wide range of natural phenomena, astronomical, physical, chemical, geological and biological, all bound together by the cosmic laws of nature. What he called *kosmische Physik* was the description of nature as a single organism, based on the view that all individual phenomena were manifestations of "the unity and generality of forces."[5] Other books used the term in a much more restricted meaning. For Alois Biegler, a gymnasium professor in Freising, Bavaria, cosmical physics comprised positional astronomy, the motions of the Earth and Kepler's laws of planetary motion.[6]

---

[4] On "Humboldtian science," see Cannon 1978, pp. 72-119, quotation on p. 77. See also Michael Bravo, "Humboldtian science," pp. 430-433 in Good 1998.

[5] Ule 1851, p 17. Otto Ule (1820-1876) gave a series of public lectures on Humboldt's *Kosmos* in 1847-1848. In 1850 he published *Das Weltall*, a work in two volumes that covered many of the same cosmophysical themes he dealt with in *Die Natur* (Ule 1850).

While Ule's book was the work of an amateur for public consumption, Johann Müller's *Lehrbuch der kosmischen Physik* of 1856 was an attempt to transform Humboldt's ideas into a branch of physical science. Although the author paid homage to *Kosmos*, modestly stating in the preface that the book was a briefer and more systematic version of Humboldt's masterpiece, in reality it was a textbook in astronomical and terrestrial physics. Müller was at the time professor of physics at the University of Freiburg in Breisgau, where his research covered areas of optics, electricity and the effects of solar radiation. In addition, he was a well-known author of textbooks in physics.[7] His book on cosmical physics was structured in four parts, with the first and largest covering classical astronomy. The second part dealt with

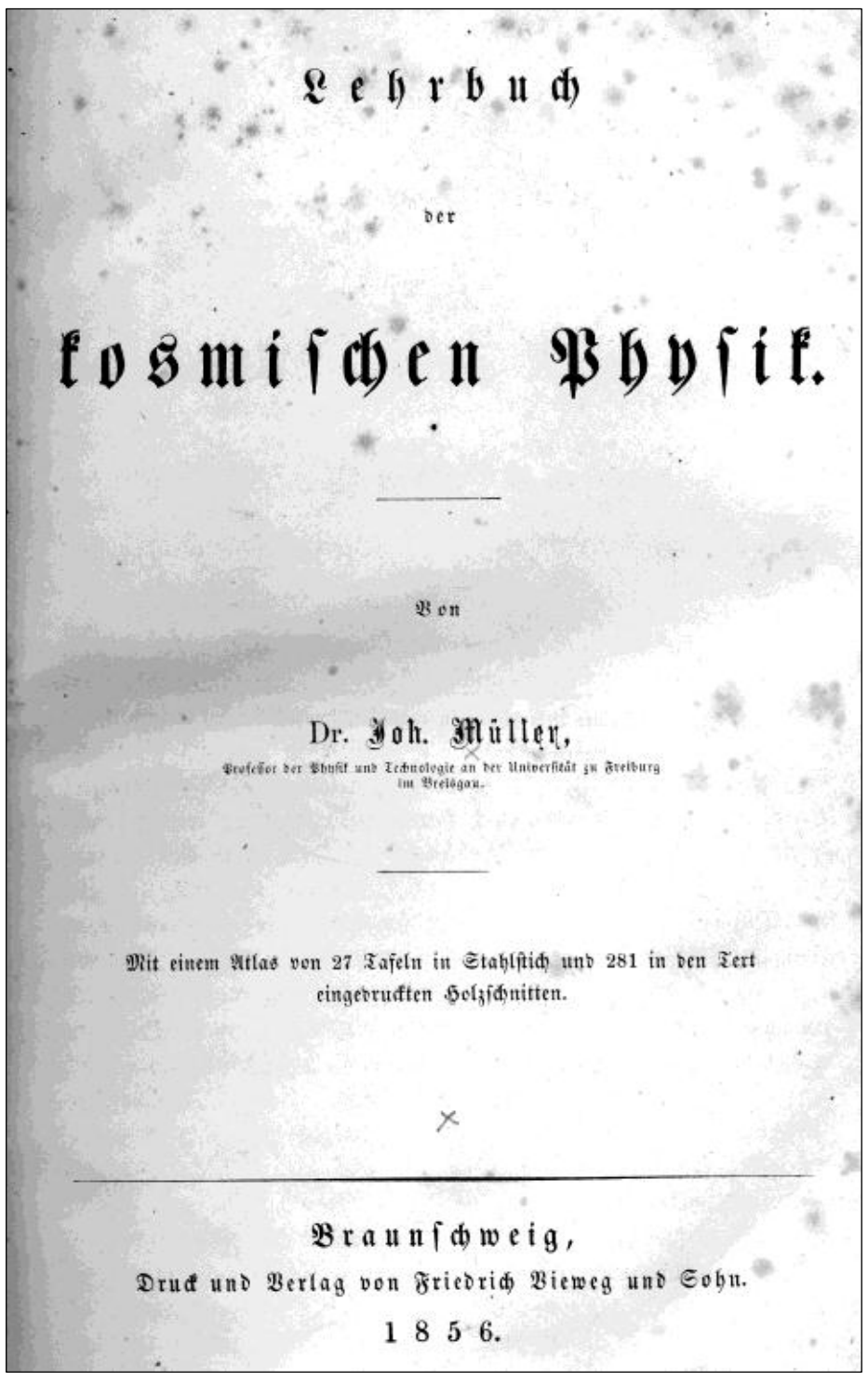
Lehrbuch

der

kosmischen Physik.

Von

Dr. Joh. Müller,

Professor der Physik und Technologie an der Universität zu Freiburg im Breisgau.

Mit einem Atlas von 27 Tafeln in Stahlstich und 281 in den Text eingedruckten Holzschnitten.

Braunschweig,

Druck und Verlag von Friedrich Vieweg und Sohn.

1856.

The first edition of Müller's textbook.

[6] Biegler 1862.

[7] On Müller as a physicist and teacher of physics, see Jungnickel and McCormmach 1986a, pp. 221-22, and Hans Kangro's entry in *Dictionary of Scientific Biography*, vol. 9.

optical phenomena in the skies and the atmosphere, and it was followed by chapters on the atmosphere and the crust of the Earth from the point of view of thermal physics. The last part mainly dealt with atmospheric electricity and terrestrial magnetism, including polar light. Müller distinguished between pure and applied physics, with the latter comprising technical physics as well as cosmical physics:

> *Cosmical physics*, which covers astronomy and meteorology, must explore large-scale natural phenomena and, in so far it is possible, base them on the laws of physics. Thus, it must demonstrate how the forces that we know from experimental physics are active throughout the entire creation; and likewise, how the same laws that we investigate in the physical cabinet govern the universe in its totality.[8]

*Lehrbuch der kosmischen Physik* became a success. During Müller's life it appeared in four, increasingly extended editions (1856, 1865, 1871 and 1875) and after his death in 1875 a final edition was published in 1894, edited and revised by the Kiel astronomer Carl F. W. Peters.[9] Whereas astrospectroscopy was not part of the first edition, written before Kirchhoff and Bunsen's invention of the spectroscope, the subject was included in the later editions, but only cursorily. Neither did Müller's book, in any of its editions, cover cosmological questions in a wider sense. For example, it did not mention the heat death of the universe, a subject much discussed in the last part of the nineteenth century.

At the turn of the century many German scientists were occupied with aspects of cosmical physics, although they rarely used the name. There were occasionally university lectures on the subject, such as in the summer semester of 1904, when the astronomer Adolf Marcuse offered lectures at Berlin University on "Astronomical Geography and Cosmical Physics."[10] However, lectures of this kind were not common and there seems to have been no chairs or institutes in Germany explicitly associated with cosmical physics.

---

[8] Müller 1856, p. 2. The textbook was accompanied by an *Atlas zum Lehrbuch der kosmischen Physik*. The various editions of Müller's book on cosmical physics also appeared as supplementary volumes to *Müller-Pouillet's Lehrbuch der Physik und Meteorologie*, a widely used textbook system based on the French physicist Claude Pouillet's *Élements de physique expérimentale et de méteorologie* from 1827. Müller first translated Pouillet's book into German in 1842. The last and final edition of *Müller-Pouillet* was published as late as 1926-1935 in five large volumes.

[9] See the review in *Nature* 50 (1894), 49-50 by William James Lockyer, the son of J. Norman Lockyer.

[10] According to *Jahresbericht der deutschen Mathematiker-Vereinigung* 1904, p. 204.

Until 1903 Müller's book was alone in defining the field, which by that time had entered the classification systems of the two leading German abstract journals in physics, the *Beiblätter zu den Annalen der Physik* and the *Fortschritte der Physik*. The *Beiblätter* started including cosmical physics in 1901, dividing it into the subclasses astrophysics and geophysics. In later volumes the field was further subdivided, so that in 1914 cosmical physics was taken to comprise the following subclasses: astrophysics (general, fixed stars); solar physics; planets and comets; geophysics (terrestrial electricity and magnetism); physics of the crust of the Earth; physics of the atmosphere, meteorology. *Science Abstracts*, the leading English-language abstract journal, was issued by the Institution of Elecrical Engineers and the Physical Society of London. Covering physics and electrical engineering, it had a narrower scope and did not include either astrophysics, geophysics or cosmical physics.

The *Fortschritte*, founded in 1845 by the Berlin Physical Society and later issued by the German Physical Society, included since its early years sections on the border regions of physics such as meteorology, physical geography and geomagnetism. In 1890 a separate volume named "cosmical physics" was published, the parts of it being astrophysics, meteorology and geophysics. According to the editor, the physicist and meteorologist Richard Assmann, although cosmical physics was not pure physics, it belonged without a doubt to the physical sciences.[11] Assmann justified the change from "physics of the Earth" to "cosmical physics" by pointing out that astrophysics did not fit under the first label.[12] In the volumes of *Fortschritte* published in the beginning of the new century, cosmical physics appeared prominently as one of the journal's seven main groups of physics, together with traditional subjects such as acoustics, electricity and magnetism, heat and optics.

---

[11] See Jungnickel and McCormmach 1986b, pp. 109-110.

[12] Preface to *Fortschritte der Physik im Jahre 1893*, vol. 49, Part 3. At the same time the previously independent sections on "terrestrial magnetism" and "atmospheric electricity" were subsumed as subsections under "geophysics" and "meteorology," respectively. The name "geophysics" replaced the older "physical geography," and the subsection on "polar light" was moved from "astrophysics" to "terrestrial magnetism." In Germany "geophysics" was regularly used in the 1880s, where the first journal devoted to the field, the physical geographer Georg Gerland's *Beiträge zur Geophysik*, was founded in 1887. In French and English literature, the term was widely adopted only after the turn of the century. See Good 2000.

Although there was no separate entry on cosmical physics in *Meyers grosses Konversations-Lexikon*, it appeared under the entry on "cosmology":

> Cosmical relations are those relations (that in numerous ways intervene in meteorology, geology, etc.) of the Earth and its inhabitants to nature at large and as a whole, to the general forces that permeate the universe, such as gravity, light, magnetism and electricity; and, moreover, to the other celestial bodies, such as the Sun, the planets and the moons, contrary to the terrestrial relations that exclusively or mostly concern the Earth. The name cosmical physics has been proposed by Joh. Müller ("Lehrbuch der kosmischen Physik," 5th edn., Braunschw. 1894) for the part of physics that relates to these phenomena.[13]

To illustrate the kind of cosmical physics entertained by some German scientists in the late nineteenth century, consider first the physicist Johann C. F. Zöllner, who in 1866 was appointed professor of "physical astronomy" (or astrophysics) at Leipzig University, first as *Extraordinarius* and from 1872 as full professor. This was the first chair of its kind. Zöllner may also have been the first to use the term "astrophysics" in its modern meaning, which he did in a paper of 1865. Zöllner did important work in stellar photometry, astrospectroscopy, cosmology and cometary theory, and he also contributed to aspects of geophysics, including an electrodynamical theory of the origin of terrestrial magnetism. He was, for all practical purposes, a cosmical physicist, although he did not identify himself as such.[14]

Another and possibly better example of a German cosmophysicist with roots in astronomy is Wilhelm Julius Förster, who since 1865 served as director of the Berlin Astronomical Observatory. As a young man Förster had met Humboldt, and the meeting with the admired scientist and explorer made a lasting impression on him. As he later recalled, it was under the impression of Humboldt's *Kosmos* that he decided to follow a career in science and to cultivate astronomy as an interdisciplinary field that related also to physics and the earth sciences.[15] Indeed, much of Förster's scientific work focused on solar and geophysical problems, including observations of geomagnetism, atmospheric electricity and the thermal behaviour of air masses. As early as 1871 he proposed to create a solar observatory in Berlin to investigate such phenomena.[16] Solar-terrestrial physics was at the heart of his research, one important part of which was the study of auroral phenomena. For example, in the 1870s he was among the first to study auroral spectra and suggest that they were due to electrical discharges originating in the Sun. He maintained his interest in the interface between meteorological and astronomical phenomena over the years, a subject he dealt with in scientific as well as popular contexts.[17]

---

[13] *Meyers grosses Konversations-Lexikon*, 5th edition of 1905-1909, and also in the 4th edition of 1888-1889. The 5th edition of the encyclopaedia included a separate entry on "cosmical meteorology," defining it as "the science of the influence of the stars on the meteorological elements and the weather." The same encyclopaedia included an entry on "geophysics" as early as 1848.

[14] On Zöllner and his work, see Kragh 2012a.

[15] Förster 1911, pp. 50-53.

Förster's interest in cosmical physics made him found in 1891 a society of "friends of astronomy and cosmical physics," the first president of which was Rudolf Lehmann-Filhés, a Berlin professor of astronomy. Förster also established an associated journal, the *Mitteilungen der Vereinigung von Freunden der Astronomie und kosmischen Physik* of which he was the chief editor. The society was organized in six sections, dealing with observations of (1) the Sun; (2) the Moon; (3) starlight and the Milky Way; (4) zodiacal light and meteors; (5) polar light, terrestrial magnetism and atmospheric electricity; and (6) clouds, halos and thunderstorms.[18] According to Förster, although cosmical physics was international in scope, some parts of it had a local character and were better organized on a national than an international basis.[19] He considered the journal to be a vehicle for the promotion of cosmical physics as he saw it – essentially geophysics and its relations to astronomy, astrophysics and meteorology – and he contributed with papers in which he emphasized the relevance of the most recent discoveries in physics, such as radioactivity and the electron structure of atoms.[20]

Although the society and its journal were intended to deal with cosmical physics in a broad sense, including its geophysical and meteorological aspects, in practice this was the case only to a limited extent. From its very beginning astronomy counted more than cosmical physics, and after the Münster astronomer Joseph Plassmann replaced Förster as editor in 1906, the predominance of astronomical subjects became even stronger. When the journal changed its name to *Die*

---

[16] See Schröder 1999 and Hedenus 2007, pp. 105-138.

[17] See, for example, Förster 1906.

[18] See *Nature* 44 (1891), 206 and 507.

[19] Förster 1903.

[20] Förster 1904. Förster 1912.

*Himmelswelt* in 1920, it reflected that the ambitions of unifying astronomical and terrestrial physics had been abandoned.

After studies at the universities of Heidelberg and Innsbruck, 25-year-old Alfred Wegener presented a doctoral dissertation at Berlin in 1905, an analysis of the Alfonsine Tables by means of modern methods of astronomical calculation. The subject was proposed by Förster, whom he knew from the people's observatory "Urania," where he had been an assistant and which Förster had helped establishing in 1888. Wegener subsequently participated in a Danish expedition to Greenland 1906-1908 and upon his return he became a lecturer (Privatdozent) in meteorology, cosmical physics and practical astronomy in Marburg. In 1924 he was appointed full professor of meteorology and geophysics at the University of Graz.

In the years from 1909 to 1915 Wegener engaged in a variety of research subjects, including upper-atmosphere meteorology, meteorites, the formation of craters on the Moon and – famously – the origin of the continents and oceans. His main output in this early phase was the impressive monograph *Thermodynamik der Atmosphäre* from 1911.[21] Wegener was particularly interested in the physical and chemical composition of the upper atmosphere, suggesting that it consisted of a mixture of hydrogen and a new gas, which he proposed to call "geocoronium." He argued that there were tiny amounts of the gas even at sea level and that it probably was a new chemical element with atomic weight smaller than the one of hydrogen. In this way he thought it was possible to explain the puzzle of the characteristic green line of wavelength 5570 Å appearing in all auroral spectra, which eventually turned out to be due to oxygen. The short-lived geocoronium hypothesis was ignored by most contemporary meteorologists.[22]

The wide range of subjects that Wegener dealt with during his stay in Marburg was in accord with the ideas of cosmical physics, for which he expressed great sympathy. In a letter to his father-in-law, the German meteorologist Wladimir Köppen, he spelled out what he hoped would be a bright future for cosmical physics in Germany:

---

[21] Wegener 1911. See also Schröder 1981.

[22] Kragh 2010. Wegener thought that geocoronium was similar to and perhaps identical with the solar "coronium" element that several astrophysicists had inferred from spectral investigations of the Sun's corona. If so, it would constitute another link between the terrestrial and cosmic regions.

> Recently I have often wondered whether "cosmical physics" might have more justification as a university subject than such a specialized subject as meteorology. In the textbooks of physics, cosmical physics figures prominently as an example. … [But] at the universities this has been abolished, for there is no longer time for such digressions. From this follows in principle the necessity that cosmical physics be taught here in special lectures complementary to experimental physics and that it should be taken into consideration to about the same extent as is the case with theoretical physics … It would be necessary that cosmical physics is represented at each university where physics is taught, and that cosmical physics is obligatory at the physics exams for high school teachers. Should thoughts like these get across, then the huge public of physics students would be won for the cause; when all the major universities have chairs in cosmical physics, some of the chairholders will take up the physics of the atmosphere as part of their work.[23]

However, Wegener's daydreams remained dreams. Although there were several courses in German universities in meteorology and allied subjects, cosmical physics did not become institutionalized in the German university system.[24]

The approach of cosmical physics was clearly visible in Wegener's famous and controversial theory of the drifting continents, which he first introduced in a paper of 1912 and three years later in an extended form in the monograph *Die Entstehung der Kontinente und Ozeane*.[25] The theory was characteristically synthetic, based on evidence from areas of science that were traditionally seen as separate. His arguments did not refer only to geology, geophysics and geodesy, but also to astronomy, paleoclimatology and zoology. It was a kind of Humboldtian science or cosmical physics that appealed to the coherence of the broader picture but invited criticism from the point of view of the special sciences. The conflict between globalism and localism was an important element in the controversy that followed in the 1920s.[26] Wegener's theory of continental drift can be considered a late example of cosmical physics, although one with its focus on the terrestrial rather than the astronomical sciences.

---

[23] Wegener 1960, pp. 72-73. The date of the letter is not given.

[24] For a list of meteorological and related courses in German-speaking universities in Europe, see *Monthly Weather Review* 34 (1906), 226-227. In 1906 a course in cosmical physics was offered only at the University of Innsbruck, where Vilhelm Trabert gave two lectures per week. At the University of Göttingen, Emil Wiechert gave courses in terrestrial magnetism, polar light and geophysics.

[25] Wegener 1912. Wegener 1915.

[26] See, for example, Le Grand 1994.

## 2. Switzerland

The basic idea of cosmical physics was to consider astronomical as well as terrestrial phenomena under the unified perspective provided by the laws of physics. According to Rudolf Wolf, a professor of astronomy in Zurich, cosmical physics took its beginning in 1852 with the discovery of a definite link between solar and terrestrial physics. In this year the sunspot cycle was correlated to the frequency of aurorae and the geomagnetic activity on Earth, independently by Wolf himself, his compatriot Jean-Alfred Gautier and Edward Sabine in England.[27]

But why restrict cosmical physics to the solar-terrestrial interaction? And why focus on physics and not, say, chemistry? Expressing the same synthetic aspirations as Müller in his *kosmische Physik*, and like him admitting inspiration from Humboldt, the Swiss physical chemist Emil Baur gave in the winter semester 1902-1903 a series of public lectures on "chemical cosmography" at the Munich Technical University.[28] With this term he meant the chemical processes in all of nature, which he divided into three groups: the chemistry of the stars; chemical transformations in the crust of the Earth; and the circulation of chemical elements in organic nature. His collection of subjects included many of those dealt with by the cosmical physicists (such as the constitution of the Sun, meteorites, comets and volcanoes), but it was even broader. Baur's chemical cosmography also covered aspects of organic nature, including biochemistry, photosynthesis and fermentation processes.

In his discussion of the temperature of the Sun, Baur introduced Max Planck's new radiation law that would soon revolutionize physics. However, to him and most of his contemporaries the law was of interest simply because it represented the spectrum of the heat radiation so accurately. He did not mention the hypothesis of energy quantization that at the time was still disputed or considered unimportant. At any rate, Baur's book on chemical cosmography was an isolated case and not an attempt to create a new framework of cosmic chemistry in the style of cosmical physics. So-called cosmochemistry would eventually be established as an extension of geochemistry, but this only happened some four decades later.[29] In almost all work within the cosmophysical tradition, whether written by authors trained in physics, astronomy, meteorology or other parts of the earth sciences, chemical perspectives were absent.

---

[27] Wolf 1892, pp. 408-421, who only referred once to cosmical physics and without referring to Müller's book on the subject.

[28] Baur 1903. After his stay in Munich, Baur was appointed professor at the Braunschweig Technical University and in 1911 he returned to Switzerland to take up a chair at the Zurich Technical University (ETH). He did important work on fuel cells and in areas of electrochemistry and photochemistry. See Treadwell 1944.

I know of only one instance of institutionalized cosmical physics in Switzerland. This was at the University of Freiburg (Fribourg), where Albert Gockel was appointed ordinary professor and director of an institute of cosmical physics in 1910. However, the Freiburg institute was tiny, not living up to its grand name: apart from Gockel, it consisted of just a single assistant. Gockel had since the beginning of the century specialized in atmospheric electricity and its relation to radioactivity, a line of research that led him to a series of balloon flights to measure the ionization of air as a function of the height. He was among the first to recognize, together with Victor Hess from Austria and Werner Kohlhörster from Germany, the penetrating extra-terrestrial radiation that in the 1920s became known as cosmic rays.[30] The early study of the mysterious cosmic rays was seen as an important part of the cosmical physics research programme.

## 3. Austria

The Austro-Hungarian Empire was the only country in which cosmical physics obtained a firm institutional setting over a longer period of time.[31] In the late nineteenth century research in meteorology and geophysics was cultivated in part at the universities and in part at the Central Bureau of Meteorology and Earth Magnetism in Vienna founded in 1851 and since 1904 renamed the Central Bureau of Meteorology and Geodynamics (Zentralanstalt für Meteorologie und Geodynamik).[32]

The leading figure in Austrian meteorology at the turn of the century was Julius Hann, who was internationally known for his dynamical theory of cyclones and anticyclones, as a long-time editor of the *Meteorologische Zeitschrift* and also as an author of pioneering textbooks in meteorology and climatology (*Handbuch der*

---

[29] On the origin and early history of cosmochemistry, see Kragh 2000 and Kragh 2001.

[30] See, e.g., Xu and Brown 1987. On Gockel's work, see Lacki 2012.

[31] On physics and chemistry, including cosmical physics, in the Austro-Hungarian Empire at the turn of the century, see Crawford 1992, pp. 82-86, and Crawford 1996, pp. 132-134.

[32] For a detailed account of the history of the Central Bureau, see Hammerl et al. 2001.

*Klimatologie*, 1883; *Lehrbuch der Meteorologie*, 1901).[33] From 1877 to 1897 he served as director of the Central Bureau and at the same time as full professor of physical geography at the University of Vienna. Not only was Hann instrumental in promoting and organizing cosmical physics, it was also due to his efforts that the field became obligatory in the university exams for would-be teachers in the secondary schools.[34] After a brief stay at the University of Graz as professor of meteorology he returned to Vienna to work as professor of cosmical physics from 1900 to 1910, a personal chair with an institute that existed by name only. The directorship of the Central Bureau had in 1897 been passed to Josef Maria Pernter (1897-1908) and subsequently to Wilhelm Trabert (1909-1915) and Felix Maria Exner (1916-1930).

The three successors of Hann had all held positions at the University of Innsbruck, which at the time was the most important institution of cosmical physics in Austria and indeed in the world. A chair was established in 1890, when Pernter, a former student of the physicist Josef Stefan and an assistant to Hann in Vienna, was appointed extraordinary professor of cosmical physics in Innsbruck.[35] Two years later he advanced to an ordinary professorship. As understood in Innsbruck and Austria generally, cosmical physics was essentially meteorology and geophysics, comprising among its subfields geomagnetism, theoretical and optical meteorology, seismology, gravitation, atmospheric electricity, climatology and oceanography. Characteristically, in a letter recommending Pernter for the chair, Stefan referred to "the extraordinary chair in cosmical physics or rather, to use another term, in meteorology and physical geography." In the appointment document from the Minister of Education, the position was described both as a chair in "cosmical physics" and in "meteorology and physical geography," the two terms apparently being used synonymously.[36]

As a professor in Innsbruck, Pernter gave lectures 1890-1892 that covered subjects such as cosmical physics, the physics of the atmosphere, astrophysics, geophysics, cosmology and meteorology. However, this broad spectrum of subjects was not reflected in the actual research of either Pernter or his students, which was restricted to classical meteorology and geophysics. Indeed, the chair was officially in the areas of physics and meteorology, or the interface between the two, whereas a candidate trained in astronomy would not qualify.[37] Pernter may have considered cosmical physics a label too broad to be of scientific significance. In 1902 he wrote a detailed book on meteorological optics, pointing out that that in the existing literature on "meteorology and cosmical physics" this topic was only treated insufficiently.[38] From the point of view of the specialist in meteorological optics, to tie it to all the other subjects of cosmical physics would only hinder progress.

---

[33] On Hann, see Süring 1922 and Shaw 1921, who called him "the chancellor of the realm of meteorology." See also the entry in *Dictionary of Scientific Biography* by Gisela Kutzbach.

[34] According to Crawford 1992, p. 85.

[35] See Oberkofler 1971 and Oberkofler et al. 1990. Crawford states wrongly that Pernter's chair in cosmical physics was at the University of Graz. There never was such a chair in Graz (Besser 2004).

[36] Stefan to the Minister of Education, 28 March 1890, quoted in Hammerl et al. 2001, p. 276. Oberkofler et al. 1990, pp. 47-53.

When Pernter replaced Hann in Vienna, the vacant Innsbruck chair in cosmical physics was filled by Paul Czermak, who had been an assistant to Ludwig Boltzmann and worked in physical meteorology in Graz. Although there was a chair in cosmical physics in Innsbruck, and formally also an institute, it did not have its own premises. Only in 1906 did the institute move into a new building, sharing it with the physics institute and some other institutes. By that time Czermak had become professor of experimental physics in Innsbruck and the chair in cosmical physics taken over by Trabert, who for a time had Albert Defant and Heinrich von Ficker as his assistants. In 1910 Exner was appointed extraordinary professor of cosmical physics in Innsbruck, where he remained for six years. After he left for Vienna, the chair was vacant for a long time, namely until 1927, when it was filled by the physicist and meteorologist Arthur Wagner.[39]

Besides Innsbruck and Vienna, chairs in cosmical physics were also established at the German University in Prague and at the University of Czernowitz, the capital of the Austro-Hungarian duchy Bokowina and today part of Ukraine (and named Chernivtsi). In Prague, Rudolf Spitaler, whose background was in astronomy and meteorology, was appointed ordinary professor in 1909. Victor Conrad, who like Spitaler had studied under Hann, was appointed professor of cosmical physics in Czernowitz in 1909, where he remained until the collapse of the Austro-Hungarian

---

[37] Oberkofler et al. 1990, p. 61.

[38] Pernter and Exner 1922, p. V. This was a second, revised edition of the 1902 book.

[39] Oberkofler 1971, pp. 144-146.

monarchy in 1918.[40] In both cases small institutes of cosmical physics were established.[41]

As the author of a comprehensive textbook in cosmical physics, Wilhelm Trabert deserves special attention. Yet another student of Hann, from 1902 to 1909 he occupied the chair in cosmical physics in Innsbruck, after which he went to Vienna to become director of the Central Bureau and university professor in the earth sciences ("Physik der Erde"). During his period in Vienna he published a textbook in meteorology and climatology and, in 1911, a *Lehrbuch der kosmischen Physik*, the third and last book with this title. Remarkably, neither in the preface nor elsewhere in the book did he mention his two predecessors, Müller and Arrhenius.[42]

Trabert's *Lehrbuch*, a work of more than 650 pages, was erudite, comprehensive and containing a fairly heavy dose of mathematics. It was also rich in historical information and with a critical, sometimes philosophical perspective that was and is rare to find in books of its kind. Apart from paying the usual homage to Humboldt, in the preface Trabert defined cosmical physics as being mainly concerned with the place of the Earth in the universe, the gravitational influences on the Earth, and the energy exchanges on the surface of the Earth and between the Earth and its surroundings in the solar system. He realized that cosmical physics might be considered just a conglomerate of already existing discipline and modestly described his book as nothing else than an attempt at "unifying in one book a series of disciplines, a loose juxtaposition of astrophysics, the physics of the solid and fluid parts of the Earth, and the physics of the atmosphere, that is, a physics that links together what in the universe is locally unified."[43]

Like Müller, Trabert started with a solid astronomical part in which he covered position astronomy and the motion of the heavenly bodies, whether planets, comets or stars. This part of about 250 pages was followed by chapters on geodesy and on solar radiation and its influence on the atmosphere. Trabert included a concise section on solar physics and spectral analysis, concluding that "the chemistry of the Earth is also the chemistry of the universe," but without elaborating on the chemical similarities.[44] After having covered standard topics such as meteorology, volcanology, seismology, climatology and the aurora borealis, he turned to terrestrial magnetism and electricity. In the final part of the book he dealt with the evolution of the Earth, the stars, the solar system and the entire universe from the perspective of thermodynamics, both the first and the second law.

---

[40] Conrad returned in 1919 to Vienna to become director of the Central Bureau and university professor. In the early 1930s he established a book series, *Ergebnisse der kosmischen Physik*, in association with the journal *Gerlands Beiträge zur Geophysik* of which he had become the editor in 1926. After the *Anschluss* in 1938, Conrad, who was a Jew and a socialist, emigrated to the United States.

[41] According to Crawford 1996, p. 133, there was also created a chair and an institute for cosmical physics in Budapest. I have been unable to verify this.

[42] The book was positively reviewed in *Nature* by Ernest Gold, director of the Meteorological Office in London, who used the opportunity to compare it with Arrhenius' "masterly treatise." *Nature* 90 (1912), 356-357.

[43] Trabert 1911, p. iv.

Trabert's textbook was unusual by dealing also with cosmology in the wider sense, the science of the universe as a whole. He discussed in some detail the problem of the anomalous motion of Mercury's perihelion that had worried astronomers since 1859 and in 1915 would be solved by Einstein by means of his new theory of gravitation. Trabert mentioned various attempts to solve the problem, favouring the possibility that Newton's law of gravitation might not be accurate on a cosmic scale. In this connection he discussed the conclusion reached by the German astronomer Hugo von Seeliger in 1895-1896, that "either Newton's law is not strictly true, or cosmic space is finite." Whereas Seeliger had not suggested the possibility of a non-Euclidean, positively curved (and therefore finite) space, Trabert found it worthy of consideration. It is conceivable, he wrote, that astronomical observations would one day demonstrate that "light, gravitation and electricity do not propagate in straight lines, but in circles."[45]

At the end of his book Trabert referred to the question, much discussed at the time, of the cosmic consequences of Clausius' law of entropy increase. Did the law imply a future "heat death" (*Wärmetod*), an irreversible cessation of all life and activity in the universe? This was also a question that Arrhenius was much concerned with (Section 7), but contrary to him Trabert saw no problem in either a materially finite universe or the heat death predicted by the second law of thermodynamics. He concluded: "The world at large develops in a definite manner, as in accordance with a certain tendency toward a goal. … We know of no force which counters these unidirectional processes. … [The tendency toward maximum entropy] is the driving agent of the development of the world, and also, as far as we know, its goal."[46]

---

[44] Ibid., p. 484.

[45] Ibid., pp. 258-259. The possibility of a closed universe was known at the time, but only seriously considered by a few physicists and astronomers (Kragh 2012b). It is remarkable that Trabert, a professor of meteorology and geophysics, belongs to the small group. On Seeliger's analysis of the infinite Newtonian universe, see Norton 1999.

As we have seen, in the period from about 1890 to 1918 several Austrian scientists had positions in cosmical physics and it is therefore tempting to infer that they did research within the new disciplinary framework. However, it is evident from their research publications[47] that they did not, or rather, that they thought of cosmical physics in such a narrow sense that the word "cosmical" had no real meaning. In reality, the Austrian speciality of cosmical physics was geophysics or meteorological physics, whereas the astronomical aspects played a secondary role or none at all. These aspects appeared prominently in Trabert's textbook, but the research activities of the Austrian cosmical physicists, Trabert included, did not correspond to the content of the book. The normal pattern in Austrian universities was that the recognition of a new discipline came through the creation of chairs. For example, this is what happened with physical chemistry at the turn of the century. In the case of cosmical physics chairs were created, but the field became recognized as a discipline in a formal sense only and almost only in Austria. It never won recognition as a proper research discipline.

## 4. Scandinavia: Norway

In the period from about 1890 to 1920, research areas related to cosmical physics were given high priority in Norway and Sweden, such as can be exemplified by the activities of three eminent Scandinavian scientists, Bjerknes, Birkeland and Arrhenius. The Norwegian physicist Vilhelm Bjerknes was a pioneer in applying the theory of hydrodynamics to meteorological phenomena in the lower atmosphere with the aim of weather forecasting.[48] As a young man he had worked as an assistant to Heinrich Hertz in Bonn and done important work on electric oscillations and resonance effects. After having been appointed professor of applied mechanics and mathematical physics at the University College in Stockholm in 1895, he gradually moved into theoretical meteorology and hydrography. In Stockholm, he became a close friend of Arrhenius and aware of the interest in cosmical physics that Arrhenius shared with other Stockholm scientists, including the meteorologist Nils Ekholm. It was in part through Ekholm that Bjerknes was drawn into meteorology, a field that became his main line of research in his later career as a professor in Oslo, Leipzig and Bergen.

---

[46] Trabert 1911, p. 645. On cosmology and thermodynamics in the period, including the heat death and proposals to avoid it, see Kragh 2008.

[47] The publications listed in *Poggendorff's Biographisch-Literarisches Handwörterbuch* include very few works of an astronomical or astrophysical nature, and none of a cosmological nature.

[48] For a detailed account of Bjerknes' work and career, see Friedman 1989.

In an important paper of 1904 in the *Meteorologische Zeitschrift*, Bjerknes outlined a programme for rational weather forecasting based on the laws of mechanics and the hydrodynamic equations of motion.[49] As he emphasized in this and other papers, meteorology was in need of a physical and mathematical foundation to complement and make sense of the many observations. This was a theme he would often return to. For example, in his inaugural lecture of 1913 as professor of geophysics in Leipzig, he reflected on the relationship between meteorology, physics and astronomy, and on the possibility of turning meteorology into an exact science:

> Just as the earth is a part of the entire cosmos, so geophysics can be conceived as a part of cosmical physics. It is divided into three branches: Physics of the atmosphere, Physics of the hydrosphere and Physics of the rigid earth; and has thus an extraordinarily extensive scope. … The physics of the atmosphere treats of the same subjects as meteorology [but] … these two sciences must not be confounded with each other. The distinction is marked for the reason that physics ranks among the so called exact sciences, while one may be tempted to cite meteorology as an example of a radically inexact science.[50]

This theme, meteorology's weak position relative to physics, was far from new. It had been discussed for decades by meteorologists worrying that their science was falling behind the rapidly developing physical sciences. Although Bjerknes' work was greatly important and instrumental in giving meteorology increased respectability in the physics community, it was not cosmical physics in the wider sense that Arrhenius and Trabert advocated. Focused on the lower level of the atmosphere, Bjerknes' meteorological programme did not need to consider the astrophysical and solar connections that other scientists regarded as crucial components in cosmical physics.

---

[49] See Gramelsberger 2009.
[50] Bjerknes 1914, p. 12.

During the first decades of the twentieth century auroral research was a Norwegian specialty and the focus of a local tradition in cosmical physics that differed substantially not only from the one cultivated by Bjerknes but also from what Austrian meteorologists and geophysicists called cosmical physics. The tradition was to a large extent created by the physicist Kristian Birkeland, since 1898 professor at the University of Oslo or what at the time was still called the Royal Frederik University.[51] Birkeland's first experimental research was in the new and exciting field of gas discharges and cathode rays that in the late 1890s led to the discovery of the electron as a subatomic particle. Combining this line of research with a growing interest in the aurora borealis, in a paper of 1896 he hypothesized that aurorae and magnetic storms were caused by high-energy charged particles (cathode rays and possibly positive ions) from the Sun being deflected and accelerated in the magnetic field of the Earth. Although the hypothesis was not entirely new, Birkeland was the first to seek evidence for it by means of experimental tests, which he did by making laboratory experiments with a magnetized sphere simulating the Earth. By bombarding his "terrella" with cathode rays he created artificial auroral displays, which he interpreted as support of his hypothesis.[52]

Birkeland's work stimulated several young Norwegian scientists to take up auroral and geomagnetic research. Carl Størmer, a newly appointed professor of pure mathematics in Oslo, specialized in calculations of the trajectories of charged particles entering the Earth's magnetic field and soon went on to develop Birkeland's theory of solar cathode rays into a more sophisticated version.[53] Another member of Norway's auroral community was Lars Vegard, a young physicist who had been Birkeland's assistant and in 1918, after Birkeland's death, succeeded him as professor of physics in Oslo. From about 1915 to 1925, Vegard extended Birkeland's ideas concerning the origin and nature of the northern light. His extension included a new picture of the upper strata of the atmosphere and the hypothesis that the Earth was enveloped by an electrified layer of nitrogen dust particles.[54]

---

[51] On the Birkeland tradition in Norwegian space- and geophysics, see Friedman 1995 and Friedman 2006. Birkeland's eventful life is described in Jago 2001 and in Egeland and Burke 2010. References to his publications can be found in these works.

[52] See Rypdal and Brundtland 1997.

[53] See the review in Nutting 1908.

[54] Vegard's theories of the upper atmosphere and the origin of the auroral spectrum are dealt with in Kragh 2009.

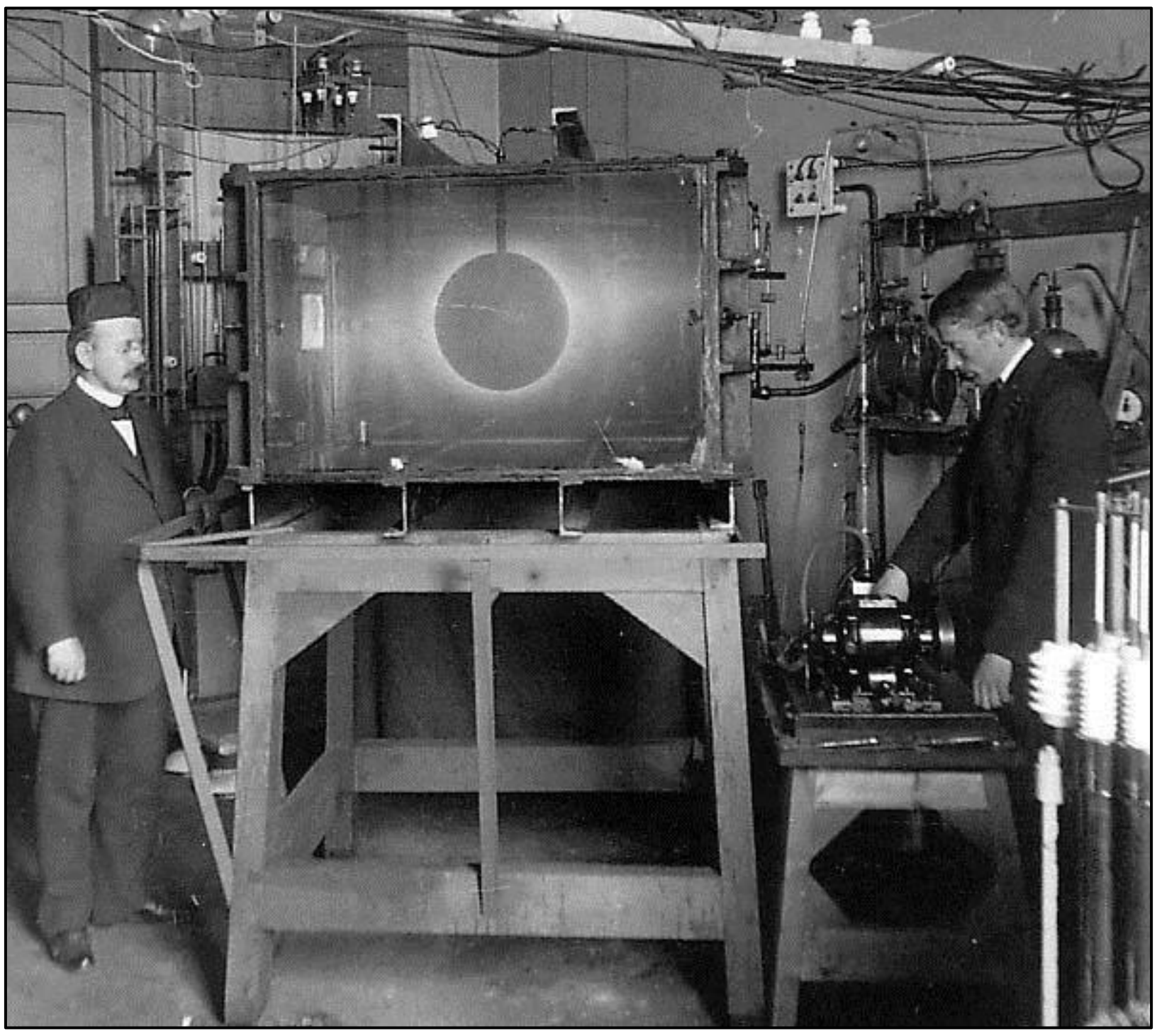

Birkeland (left) and his assistant Karl Devik showing discharge phenomena with the large terrella in 1913.

Although Birkeland rarely used the term "cosmical physics," much of his work can be classified as such. Solar-terrestrial physics was at the heart of his wide-ranging research programme, but it was only one part of it.[55] It also included the weather and climate of the Earth, geomagnetic storms, cometary tails, planetary rings, the zodiacal light, the formation of planets and other cosmic phenomena. For example, he simulated in his laboratory in Oslo the rings of Saturn and suggested that the planets in the solar system were formed by electrified matter ejected from the Sun.[56] The Sun itself might be a radioactive machine:

> The disintegration theory, which has proved of the greatest value in the explanation of the radio-active phenomena, may possibly also afford sufficient explanation as to the

---

[55] Solar-terrestrial physics includes the impact of solar activity on terrestrial phenomena such as geomagnetic storms and the aurora borealis. For a summary history, see E. W. Clive, "Solar-terrestrial relations," pp. 777-787 in Good 1998.

[56] Birkeland 1913a. Egeland and Burke 2010. Birkeland compared his cosmological scenario with the one of Arrhenius, stating that his own view was superior to Arrhenius' because it was based on "experimental analogies" (p. 563). Although he recognized the similarity of his own cosmogony to the one expounded by Arrhenius, he denied that the radiation pressure played the dominant role that Arrhenius ascribed to it (Birkeland 1913b, pp. 20-21).

> origin of the sun's heat. The energy of the corpuscular precipitation that takes place in the polar regions during magnetic storms seems indeed to indicate a disintegration process in the sun of such magnitude, that it may possibly clear up this most important question in solar physics.[57]

To the mind of Birkeland, the aurora was of cosmological significance and not merely a phenomenon in the upper atmosphere. In his grand theory of the solar system, electromagnetic forces were no less important than the gravitational force of traditional cosmogony. Not only was the Sun a generator of electrons and ions, the same was the case with the other stars, which led Birkeland to speculate that interstellar space was filled with a tenuous ionized gas – a plasma. Indeed, his theory was later seen as a precursor of the so-called plasma universe, although the term "plasma" had not been coined yet.[58]

According to Birkeland, "the greatest amount of matter in the universe is not to be found in stars or in nebulæ, but in interstellar airless space, which may be supposed completely filled with flying corpuscles of every nature – electrons, atoms, molecules or coarser particles, electrified or not, representing all the chemical elements."[59] Some of these particles were held to be due to a hypothetical kind of radioactivity for which he coined the name "electro-radioactivity." The mechanism led him to speculate that "the gases of which our atmosphere and the oceans were originally composed, were produced by a degradation of elements in the course of the same electric disintegration of terrestrial matter which finally led to the creation of our moon."[60] He used his hypothesis – claimed to be supported by experiments – to explain such diverse phenomena as the ice ages, the geological periods, Saturn's rings, the eleven-year period of sunspots, cepheids and other periodically varying stars, and the spiral nebulae. Nothing less than a cosmic theory of everything – a "mechanism of the universe," he called it – it further led to the conclusion that "the worlds [stars and nebulae] in the universe must be infinite in number."[61]

---

[57] Birkeland 1908, p. IV. The idea that solar energy is in part of radioactive origin was popular in the first decade of the twentieth century (Kragh 2007).

[58] The name "plasma" was introduced by the American physical chemist Irving Langmuir in 1928. The Swedish physicist Hannes Alfvén, a Nobel laureate of 1970, developed in the 1960s a cosmological theory based on plasma physics which, he claimed, explained the expansion of the universe without assuming a violent beginning in a big bang. He considered Birkeland to be the first cosmic plasma physicist and his own work to be in the tradition of Birkeland. For a historical and philosophical perspective on Alfvén's plasma theory, see Brush 1990.

[59] Birkeland 1913b, p. 20. The hypothesis may be considered an anticipation of the notion of "dark matter" that later became important in cosmology.

[60] Ibid., p. 12. The idea of a general degradation of matter was popular at the time, when it was often assumed that all chemical elements were radioactive.

The auroral and cosmical research done by Birkeland and his associates attracted public attention but only limited scientific recognition. Astronomers ignored his cosmological ideas. Many scientists considered his theories to be speculative and based on too easy analogies between laboratory experiments and natural phenomena. On the other hand, it was just the experimental basis of Birkeland's research that impressed other scientists. In a public lecture of 1909 the American meteorologist Cleveland Abbe praised the work of Birkeland and Størmer on the nature of the aurora and its bearing on the upper atmosphere. He thought that the theory of the two Norwegians promised "to carry us from the firm ground of experimental laboratory physics over into the equally firm, but unexplored region of mathematical cosmical physics."[62]

**5. Scandinavia: Sweden**

Science in Sweden at the turn of the century was largely located at the two old universities in Uppsala and Lund, in the Swedish Royal Academy of Sciences in Stockholm, and at the private University College (Stockholm Högskola, later the University of Stockholm) founded in 1878. As noted by Swedish historians, the attitude to and style of science adopted by scientists at the Stockholm College differed markedly from the attitude and style predominant in nearby Uppsala and also in Lund in the southern part of the country.[63] Not only were the Stockholm scientists more open to the public and the media, they also tended to emphasize interdisciplinary aspects and research projects of a broad kind that included field work and, for example, arctic explorations. The Uppsala school of science, on the other hand, focused on laboratory work within the established disciplines. It put high value on careful measurements such as epitomized by precision spectroscopy, a line of research that was considered the very essence of modern science. Broadly speaking, while Uppsala scientists were specialists and restrictionists, Stockholm scientists were generalists and expansionists.[64]

---

[61] Ibid., p. 20. Like Arrhenius, Birkeland argued against the standard view of a limited stellar universe that at the time was held by a majority of astronomers, according to whom the material universe was largely identical to the Milky Way system.
[62] Abbe 1911, p. 27.
[63] Widmalm 2001. Holmberg 2007, pp. 78-80.

It is no coincidence, then, that cosmical physics flourished in Stockholm, but not elsewhere in Sweden. On the initiative of Arrhenius, in 1891 a Physics Society (Fysiska Sällskapet) was established at the Stockholm College, with the term "physics" being taken in the broad sense of the physical sciences, including fields such as meteorology, geophysics, astrophysics, oceanography and physical chemistry.[65] During the next two decades the Society served as an informal centre of Swedish cosmical physics. Apart from the towering figure of Arrhenius, leading members of the Physics Society comprised Vilhelm Bjerknes, the geochemist Arvid Högbom and the meteorologist Nils Ekholm, who had led an expedition to Spitsbergen during the International Polar Year of 1882-1883. At the time of the foundation of the Society, Arrhenius was already familiar with the name and concept of cosmical physics, a term he first used in a paper of 1888 in which he examined the influence of ultraviolet light on the electrical conductivity of the atmosphere.[66] The paper was published in the Austrian-German journal *Meteorologische Zeitschrift* edited by Julius Hann.

Many of the papers that Arrhenius published in the 1890s were of a physical-meteorological nature, in so far that they dealt with changes in the atmosphere caused by cosmic and other sources. The best known of them is undoubtedly his 1896 model of how variations in the content of carbon dioxide in the atmosphere cause climatic changes including the ice ages. Today known as a classic of the greenhouse effect (a name Arrhenius did not use), this work arose from discussions in the Physics Society and rested on Högbom's ideas of the carbon cycle in nature.[67] Other of Arrhenius' papers from the same period were more clearly in the tradition of cosmical physics, as they related terrestrial phenomena to solar, lunar and other extra-terrestrial causes. Assuming that the Moon was negatively electrified, he and Ekholm studied the influence of our satellite on the electric state of the Earth's atmosphere and also on thunderstorms and the aurora borealis. Without Ekholm, he even tried to demonstrate the physiological and psychological influence of the Moon on humans, as were he a modern meteorological astrologist.

---

[64] For the terms "restrictionism" and "expansionism," see Graham 1972.

[65] Crawford 1996, pp. 120-123, 135-137.

[66] Arrhenius 1888. Arrhenius had stayed for a period in Graz, studying under Boltzmann.

[67] Arrhenius 1896. For context and analysis, see Crawford 1996, pp. 145-155, and Crawford 1997. Arrhenius did not relate the increased temperature specifically to industrially produced carbon dioxide, but explained climatic changes in the geological past as primarily due to volcanic activity.

Among the friends of Arrhenius and members of the Physics Society in Stockholm was also Otto Pettersson, trained as a chemist and today recognized as a pioneer of physical oceanography. Much in the style of Arrhenius, he investigated the relationship between sunspots and climatic changes, and between lunar periods and solar activity. Pettersson called these investigations studies in cosmical physics.[68]

In a work of 1900 Arrhenius suggested an explanation of the aurora, and also of the tails of comets, in terms of the Sun's radiation pressure. The existence of a feeble light or radiation pressure was widely assumed but had not yet been established experimentally. This was first done by the Russian physicist Peter Lebedev in 1901, whose discovery was independently confirmed by the American astrophysicists Ernest F. Nichols and Gordon F. Hull two years later. According to Arrhenius' hypothesis, aurorae were due to very fine, electrically charged particles expelled from the Sun and carried toward the atmosphere of the Earth by the radiation pressure. When the dust particles reached the outer atmosphere they would be discharged and cause the colourful displays characteristic of the aurorae.[69]

Arrhenius used the same kind of mechanism in a theoretical study of the solar corona published in the *Astrophysical Journal* where he suggested that the dust particles were molten iron drops of varying size and mass. The lighter drops, he said, would be "driven away by the pressure of radiation, so that just the drops which, so to say, swim under the equal influence of gravitation and pressure of radiation will accumulate in the corona."[70] This was his favourite mechanism not only for the Sun and solar-terrestrial phenomena, but also for the tails of comets and indeed for the cosmos as a whole. As he saw it, the electrified dust particles emanating from the stars suggested an explanation of the mysterious nebulae and secured a universe of the kind he wanted, namely, in an everlasting state of dynamical equilibrium.

During the academic year of 1899-1900 Arrhenius gave a series of lectures on cosmical physics at the Stockholm College, and shortly thereafter he started writing a textbook on the subject. His *Lehrbuch der kosmischen Physik*, written on the request by

---

[68] E.g., Pettersson 1914.

[69] Arrhenius 1900. John Cox, a physicist at McGill University, Montreal, gave a detailed review of Arrhenius' theory, which he found to be fascinating and "at least plausible" (Cox 1902).

[70] Arrhenius 1904, p. 228. See also the summary in Arrhenius 1903, pp. 150-156, 920-925.

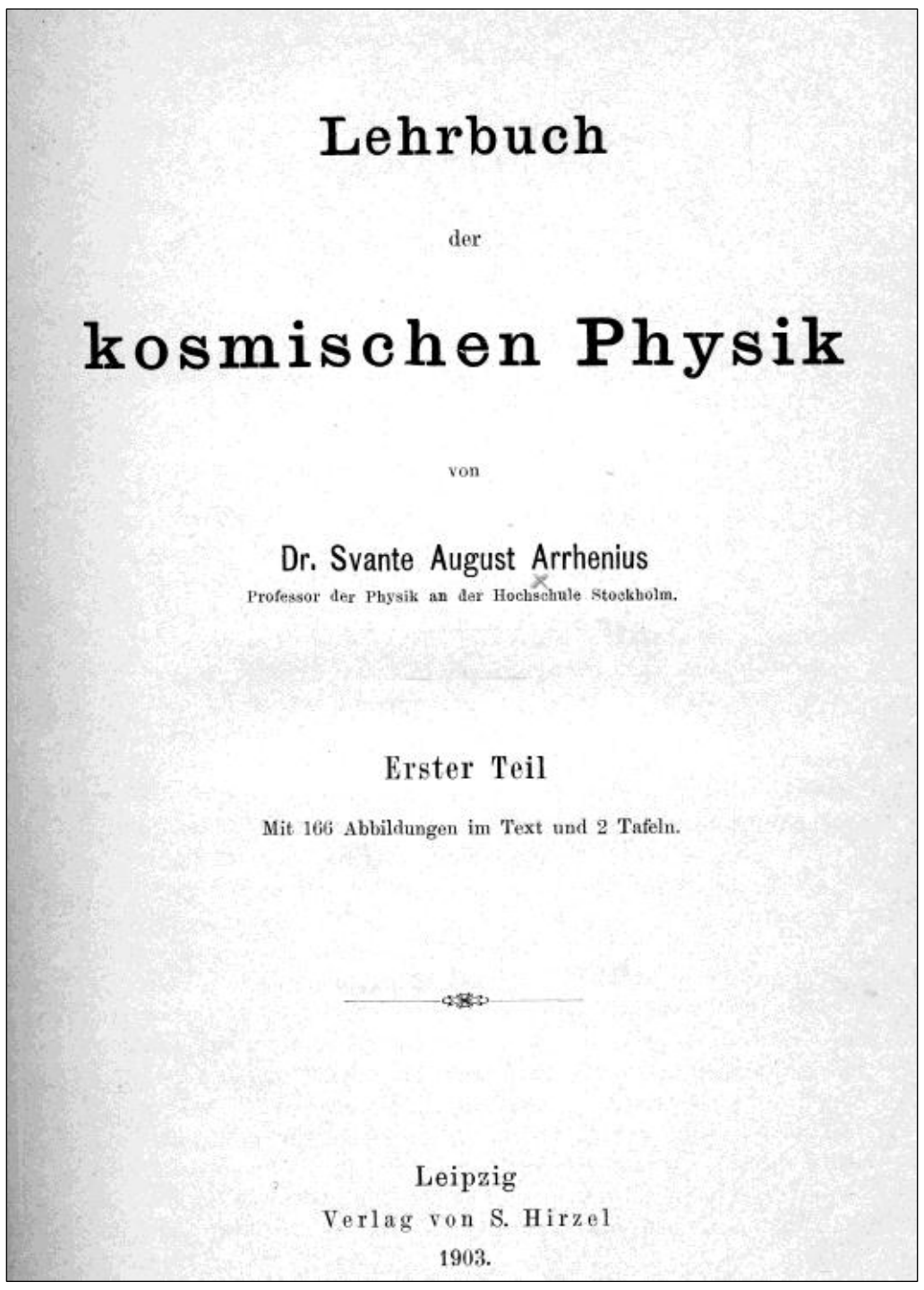

Lehrbuch

der

kosmischen Physik

von

Dr. Svante August Arrhenius

Professor der Physik an der Hochschule Stockholm.

Erster Teil

Mit 166 Abbildungen im Text und 2 Tafeln.

Leipzig

Verlag von S. Hirzel

1903.

Arrhenius' textbook in cosmical physics.

the Leipzig publishing house Hirzel and published in 1903, was a treatise of more than thousand pages organized in two richly illustrated volumes. Like Müller's earlier textbook, Arrhenius started with chapters on astronomy, but with the significant difference that it was under the heading "Celestial Physics." His focus was astrophysical rather than astronomical in the traditional sense, and with spectral analysis appearing prominently and early on. Arrhenius then went on to the physics of the Earth, covering the main branches of geophysics including volcanology, oceanography, hydrography and the age of the Earth. In the subsequent chapters on the physics of the atmosphere he dealt with, for example, meteorological optics, atmospheric electricity, the aurora and geomagnetism. A thirty-page chapter on circulations in the atmosphere was written by Johan Sandström, an assistant of Bjerknes.

Given that the fifth and much enlarged edition of Müller's similarly titled textbook had appeared less than a decade earlier, it is noteworthy that Arrhenius failed to mention either the book or its author. It is hard to avoid the impression that he wanted to appropriate the field of cosmical physics as his own invention and for this reason ignored his predecessors. Without offering a definition of cosmical physics, in the preface he explained how he conceived the new interdisciplinary research programme:

> In recent time few branches of science have made such great progress as cosmical physics. … There is almost no part of this extraordinarily versatile science … which in this respect [accumulation of data and new insights] has not presently lived through a period of blossoming. … In the treatment of this theme I have tried to avoid questions that relate purely to astronomy, hydrography, geology and meteorology; and, as far as possible, I have dealt only with such problems that have intimate connections to physics and chemistry.[71]

Although the *Lehrbuch* was in many respects an encyclopaedic compilation of data rather than a work pointing toward future research, Arrhenius made an effort to keep it as updated as possible. For example, he discussed Balmer's and Rydberg's formulae for the spectral lines of hydrogen that a decade later would be explained by Niels Bohr's atomic theory. Also worth noting is his use of Planck's radiation formula, or what he called the Wien-Planck formula, for blackbody radiation introduced by Max Planck less than three years earlier. As mentioned, the formula also appeared in Baur's chemical cosmography of the same year.

Arrhenius' *Lehrbuch* was impressive but not very successful. It was translated into Russian, but neither into English nor French, and it only appeared in the single edition of 1903. Carl Snyder, an American economist and science writer, praised "the text-book of cosmical physics from the pen of Arrhenius" as an "original and stimulative work [in which] we see for the first time in its entirety the cycle of the cosmic machine."[72] But Snyder was an amateur and an exception. While the 1894 edition of Müller's textbook in cosmical physics was extensively reviewed in *Nature*, the journal chose to ignore Arrhenius' book. Perhaps the reason was, as suggested by a historian, that it was "a textbook for a discipline that did not exist."[73] One of the problems with the book, and with the cosmical physics project generally, was its breadth: specialists in one field of research, for example meteorology or astrophysics, would find the book superficial in that particular field, whereas they would pay little attention to the other parts of the book and fail to appreciate its synthetic aim. Thus, a review in *Science*, written by an American meteorologist, dealt only with the book's meteorological chapters, which the reviewer found to be disappointing.[74] Also the German climatologist Gustav Hellmann, professor at the Prussian Meteorological Institute in Berlin, expressed reservations with regard to Arrhenius' decision to leave out questions of a purely meteorological nature. Yet, Hellmann thought the book lived up to the aim of cosmical physics, which he formulated as "the integration into one closed system of all applications of the laws of physics to problems of the heavens and the Earth."[75]

---

[71] Arrhenius 1903, pp. iii-iv. Arrhenius seems to have used the term "cosmical physics" only in the book's title and preface, whereas he did not refer to it in the main text.

[72] Snyder 1907, p. 12.

[73] Fleming 1998, p. 80.

## 6. Nobel contexts

In the early years of the Nobel institution, it was agreed that whereas astrophysics was eligible, pure astronomy was not. Also meteorology and allied sciences, including cosmical physics, were initially seen as unproblematic in the context of Nobel physics. For example, Hann was nominated thrice for the Nobel Prize in physics, in 1906, 1911 and 1913, and in none of the cases were the nominations disallowed with the argument that his work did not belong to physics.[76] Hann did not receive the coveted prize.

At a meeting of 1902 in the Physics Committee, the professor of meteorology Hugo Hildebrandsson called for strengthening the expertise in cosmical physics, an area of research he had a great deal of sympathy for.[77] In 1915 and again in 1916 and 1917, Birkeland and Størmer were nominated for a Nobel Prize in physics, although with no success. The nominator was the Swedish physicist Vilhelm Carlheim-Gyllensköld, since 1910 a member of the physics Nobel Committee, who was a specialist in geomagnetism and in the 1880s had made observations of the aurora from Spitsbergen. By proposing Birkeland and Størmer he hoped to establish a department of cosmical physics at the Nobel Institute in Stockholm, such as he proposed in 1917. Although Carlheim-Gyllensköld's proposal received some backing from other members of the Nobel Committee, it was sabotaged by Arrhenius, the leading cosmical physicist in Sweden. Arrhenius' reason was not dislike of cosmical physics, of course, but part of his plan to keep astronomy and astrophysics out of the Nobel physics institution.[78] Moreover, he may have been jealous of Birkeland for his pioneering and widely acclaimed research on aurorae, a field that Arrhenius considered himself a specialist in.[79]

---

[74] *Science* 18 (1903), 498-499, by "R. DeC. W." On the other hand, the American physicist Carl Barus, professor at Brown University, found the book to be valuable and praised it for being "an exhaustive treatise on meteorology" (*Journal of Physical Chemistry* 7, 1902, 467-469).
[75] *Meteorologische Zeitschrift* 20 (1903), 527-528.
[76] In 1906 Hann was nominated for the physics Nobel Prize by the German meteorologist and physicist Wilhelm von Bezold, in 1911 by the Austrian geologist Eduard Suess, and in 1913 by the Swedish meteorologist and close friend of Arrhenius, Nils Ekholm (Crawford 2002).
[77] Friedman 1981, p. 793.

When Arrhenius resigned his physics professorship at the Stockholm College to become head of the new Nobel Institute for Physical Chemistry, the influential Swedish mathematician Gösta Mittag-Leffler wanted Birkeland to replace him and, if possible, to become a member of the Nobel physics committee. Mittag-Leffler disliked Arrhenius, whom he considered a good chemist turned into a poor physicist. Moreover, he was involved in a long-time feud with Arrhenius over the physics Nobel Prize that centered on Mittag-Leffler's campaign to award Poincaré the prize. The campaign failed, primarily because of Arrhenius' determined opposition. Mittag-Leffler persuaded Birkeland to come to Stockholm in May 1905 to discuss the position at the College with, among others, Arrhenius and Bjerknes, but in the end Birkeland decided not to apply for the position.[80] There is little doubt that Arrhenius considered him a competitor and used his influence to avoid the Norwegian physicist as his successor.

In the 1920s several physicists within the Nobel committee campaigned against meteorology in particular and geophysics in general. One of the victims was Bjerknes, whose many nominations – between 1923 and 1945 he was nominated twelve times – never led to a prize. As a result of the campaign, geophysics, meteorology and other branches of cosmical physics were *de facto* excluded from the Nobel physics committee.

---

[78] See Friedman 1981 and Friedman 2001, pp. 147-149. Størmer was also nominated for a physics prize in 1925, 1927 and 1934, and Birkeland for a chemistry prize in the years 1907, 1909, 1912 and 1913 (Crawford 2002). The nominations for a chemistry prize had nothing to do with his work on auroral physics, but were for his co-invention with Samuel Eyde for a process of manufacturing saltpetre (calcium nitrate) from atmospheric nitrogen.

[79] See Crawford 1996, p. 161, who characterizes Arrhenius' attitude to Birkeland as "childish behavior." In Arrhenius' voluminous textbook on cosmical physics, he only referred once to Birkeland (Arrhenius 1903, p. 970), and then in a chapter on geomagnetism. Birkeland's name did not appear in his chapter on the aurora borealis.

[80] Stubhaug 2010, p. 482 and p. 506. On the conflict over Poincaré as a Nobel laureate, see Friedmann 2001, pp. 49-53.

## 7. Arrhenius' cosmological speculations

Cosmology and cosmogony in the wider sense of the terms – the structure and origin of the universe at large – had no place in the large majority of works on cosmical physics that focused on the Earth and its relations to the Sun or the solar system. However, in a few cases, such as in Birkeland, Arrhenius and Trabert, a cosmological perspective did enter, if in the case of Trabert only in a limited way. Birkeland and Arrhenius were in this respect exceptions, since both of them extrapolated their ideas of cosmical physics into elaborate cosmological theories. The main difference between the two Scandinavian cosmical physicists was that Arrhenius' theory went even farther by including the origin of life, and also that it (in part for this reason) attracted more attention than Birkeland's theory. In any case, both cosmological theories were unorthodox as seen from the point of view of the astronomers. In the history of cosmology they remain parenthetical and outside the main line of development.

Arrhenius' *Lehrbuch* of 1903 included a chapter on "Cosmogony" in which he objected to the idea of a cosmic heat death based on the second law of thermodynamics. The universal validity of Clausius' law of entropy increase had been discussed at the meetings of the Physics Society, where Ekholm were among those who had questioned it. According to Ekholm, the universe was filled with an amount of dark matter hugely greater than the amount of luminating matter, which he argued was incompatible with the "dissociation hypothesis."[81] By postulating an exchange of matter and energy between the hot stars and the supposedly cold nebulae, Arrhenius thought to have found a mechanism that would stabilize the universe and keep the heat death in check eternally. The notion of a finite-age universe, he wrote, "is hard to bring into agreement with the indestructability of energy and matter."[82] Indeed, as he made clear in his later writings, he found the notion to be non-scientific as it was impossible to comprehend creation or destruction of matter and energy in an absolute sense.

---

[81] Ekholm 1902.

[82] Arrhenius 1903, p. 232. Crawford 1996, p. 162 wrongly states that cosmogony did not form part of the *Lehrbuch*. On Arrhenius' cosmological views, see Amelin 1993 and Kragh 2008, pp. 165-173. The idea of stellar collisions as a counterentropic mechanism had earlier been suggested by the Scottish amateur geologist and astronomer James Croll in a book of 1889. Best known for his astronomical theory of the ice ages, Croll worked extensively in areas that belonged to cosmical physics.

In papers and books between 1903 and 1913 Arrhenius developed his ideas into a qualitative cosmological theory of a universe infinite in both space and time. He argued for what later became known as the cosmological principle, namely, that the universe is uniformly populated throughout with stars and nebulae, a claim that he justified methodologically rather than observationally. Since he also insisted that the universe is infinitely large, he was forced to accept a universe containing an infinity of gravitationally interacting stars, including the conceptual problems that such a picture implies. In 1896 Hugo von Seeliger had demonstrated that such a model of the universe could not be brought into agreement with Newton's law of gravitation without leading to physical absurdities. Arrhenius, aware of Seeliger's objection but without properly understanding it, denied that there was a problem with the infinite Newtonian universe.[83] As mentioned, nine years later another cosmical physicist, Vilhelm Trabert, would reach a different conclusion.

Arrhenius' best-selling *Worlds in the Making* was in some respects a popular and much reduced version of his *Lehrbuch*, only was it more speculative and focused on issues of cosmology. He repeated his objections to the cosmic validity of the second law of thermodynamics, maintaining that the infinite universe was analogous to a vast thermal machine operating with two heat reservoirs. In the hot stars the entropy would increase, but the growth in entropy would be compensated by its decrease in the much cooler nebulae. What Arrhenius has in mind, was a kind of Maxwell's demon on a cosmic scale. He did not attempt to explain the origin of stars and nebulae, nor did he think that such explanation was possible. His hypothesis was solely concerned with "how nebulæ may originate from suns and suns from nebulæ."[84] What mattered was that the combined mechanisms of radiation pressure and collisions between nebulae would allow that "the evolution of the world can continue in an eternal cycle, in which there is neither beginning nor end, and in which life may exist and continue forever and undiminished."[85]

---

[83] Arrhenius 1909, where he argued for an infinite universe. For a careful analysis of Seeliger's paradox and responses to it, see Norton 1999. The German physicist Caspar Isenkrahe suggested that the debate concerning the infinite universe and the validity of the entropy law was part of "cosmical physics" but used the term as synonymous with "cosmological physics" (Isenkrahe 1915).

[84] Arrhenius 1908, p. xiii.

[85] Arrhenius 1908, p. 211. The book appeared in Swedish in 1907 and was translated into German the same year. Other translations were in French, Russian, Italian, Finnish, Hungarian, Czech and Chinese.

This was far from an unorthodox idea at a time when the notion of the universe having a beginning was widely considered weird, nearly incomprehensible. Eternally cyclic conceptions of the universe were popular in the fin-de-siècle period, and Arrhenius' was only one among many.[86] On the other hand, astronomers generally avoided issues of a cosmological nature, including the question of the heat death.

What made Arrhenius' version stand apart from most other conceptions was not only that its author was a distinguished scientist, but also that he expanded his arguments for the eternal universe to the eternity of life in it. First in a little known paper of 1903 and then in greater detail in *Worlds in the Making*, he proposed a theory of "panspermia" according to which primitive life in the form of bacteria was propelled through space by the pressure of stellar radiation.[87] Arrhenius did not explain where these seeds of life came from and thus offered no explanation of the origin of life in an absolute sense. Nor was this what he aimed at, for he considered life to be an elementary quality that had always existed, just like energy and matter.[88] It had no origin in time.

The cosmological ideas of the Swedish chemist and physicist were widely discussed and attracted much popular interest. In a book of 1907, Snyder described and fully endorsed the views of the Swedish genius. He quoted a private letter, in which Arrhenius summarized his position:

> Formerly we knew only about the Newtonian force of gravitation, and therefore cherished the idea that in the end everything would clump together. But now we know of the presence of the pressure of radiation which may balance the tendency of congregation. … I therefore adhere to the idea of an oscillation of matter because it is impossible for me to understand a beginning or an end of the system of matter that we observe. If there were an end, with complete rest, the condition would have been reached in the infinity of time which lies behind us, and there would be nothing left in

---

[86] See Kragh 2008.

[87] Arrhenius 1908, pp. 220-230.

[88] Kamminga 1982. Some of the later theories of a "biological universe," such as due to Fred Hoyle and Chandra Wickramasinghe in the 1980s, can be considered successors of Arrhenius' ideas. For the revival of the panspermia hypothesis, see Dick 1996, pp. 367-377.

> the world for us to observe. Therefore also the second law of thermo-dynamics cannot be perfectly true as it is formulated now.[89]

Although Arrhenius' ideas were well known also in the scientific community, they failed to win approval and died a quiet death after about 1920. Arrhenius himself stuck to them. His last article, written at the time of his death, was about the panspermia hypothesis.

**8. Great Britain**

A form of cosmical physics was promoted by a few British scientists in the Victorian era. The term was known to both scientists and the public, but not used widely or with a specific meaning. One of the period's many popular science periodicals was the short-lived *Geological and Natural History Repertory* published in the mid-1860s. According to the impressive subtitle of the periodical, it was *An Illustrated Popular Weekly Magazine of Geology, Palæontology, Mineralogy, Natural-History, Terrestrial and Cosmical Physics*.

The physicist Balfour Stewart, who from 1870 to his death in 1887 served as Professor of Natural Philosophy at Owens College, Manchester, was an advocate of the "cosmical meteorology" approach according to which there was a close connection between solar physics and meteorological phenomena. Together with the astronomer J. Norman Lockyer, in 1868 he sketched a research project aiming at studying the interrelation of planetary motions, sunspot cycles and terrestrial weather patterns.[90] In 1870 the Manchester chair in natural philosophy was split into two, one of which was in mechanics and mathematical physics. The subject areas of the other, which was Stewart's responsibility, were energy physics and "cosmical physics including … astronomy, meteorology and terrestrial magnetism."[91]

Indeed, in his inaugural address at Owens College, Stewart reviewed "Recent Advances in Cosmical Physics." These advances, he said, "tend to indicate the probable union of the various branches of observational inquiry into one great cosmical research, and point to the wisdom of a very close union between the workers in the cognate fields of meteorology, terrestrial magnetism, and celestial physics."[92] Stewart argued that there was a need to strengthen physical or cosmical meteorology and generally to bring the earth sciences into closer contact to astronomy. He ended his address with a plea for organizing such a research programme on a national scale:

> Referring more particularly to Cosmical Physics, I feel convinced that meteorology should be pursued in connection with terrestrial magnetism and solar observations; and were a well-considered scheme for solving this great problem introduced, I am sure that scientific institutions and individuals throughout the country would do all that they possibly could to promote this most important branch of physical research.[93]

---

[89] Snyder 1907, p. 461, who states no date of the letter. Snyder wrote a glowing review of the German edition of Arrhenius' "epochal volume" in the *New York Times*, 6 July 1907, where it appeared under the misleading title "New Epic of Creation."

[90] Lockyer and Stewart 1868. The Stewart-Lockyer collaboration is detailed in Gooday 2004.

[91] See Kargon 1977, p. 215.

A few years later Stewart compared solar physics, terrestrial magnetism and meteorology to three corners of a triangle that are bound together: "Thus the three things hang together, and scientific prudence points to the desirability of their being studied together as a whole."[94] What he had in mind was a centralized research institution of some sort.

However, his call for support to or institutionalization of meteorological or cosmical physics fell on deaf ears. Stewart, Norman Lockyer, Arthur Schuster, William Huggins and other British scientists continued to work on aspects on cosmical physics, and especially on solar physics, but no single "well-considered scheme" emerged from their efforts.[95] The most distinguished of the Victorian physicists, Wiliam Thomson (since 1892 Lord Kelvin), may well be considered to belong to the group, as he worked on and off on cosmical physics for half a century. When Joseph Larmor edited volume 5 of Thomson's *Mathematical and Physical Papers*, he arranged 38 papers in a section labelled "Cosmical and Geological Physics." They included topics such as the age and rigidity of the Earth, its thermal history, the

---

[92] Stewart 1870, p. 500. See also Stewart 1869, a rather pessimistic evaluation of the scientific level of meteorology. For the Manchester interest in cosmical meteorology and physics, see Gooday 2007.

[93] Stewart 1870, p. 501.

[94] Stewart 1877, p. 47. J. Norman Lockyer agreed that the meteorology of the future would have to be "a physical science, and not a mere collection of weather statistics" (Lockyer 1872, p. 101). See also Anderson 2005.

[95] On solar physics in the British Victorian tradition, see Schaffer 1995. For general reviews of the rise of solar physics and astrophysics, see Hufbauer 1993, pp. 42-80 and Eisberg 2003.

duration of the Sun's heat, magnetic storms, tides, and the formation of stars and nebulae.[96]

The 1860s witnessed a dramatic change from classical astronomy to astrophysics or "astronomical physics," a change that sometimes involved also a reconsideration of the connection to the meteorological and geophysical sciences. In her *History of Astronomy during the Nineteenth Century*, the astronomy writer Agnes Clerke pointed out that traditionally the science of astronomy had been quite different from the terrestrial sciences. "It was enough that she [astronomy] possessed the telescope and the calculus," she said.[97] But now:

> She is concerned with what the geologist, the meteorologist, even the biologist, has to say; she can afford to close her ears to no new truth of the physical order. Her position of lofty isolation has been exchanged for one of community and aid. The astronomer has become, in the highest sense of the term, a physicist; while the physicist is bound to be something of an astronomer. This, then, is what is designed to be conveyed by the "foundation of astronomical or cosmical physics."[98]

What Clerke envisioned was a science that was as universal as nature was universal, a result of nature being "the visible reflection of the invisible highest Unity." The religious sentiment behind Clerke's vision can be found also in some other British scientists, including Stewart.[99] However, it was foreign to most advocates of cosmical physics. At any rate, Clerke's support of "astronomical or cosmical physics" was in reality of astrophysics and only included meteorology and geophysics on a rhetorical level. She had nothing to say in her book about these topics.

The more comprehensive meaning of cosmical physics was the one that appeared in the 1911 edition of *Encyclopaedia Britannica* and according to which the field differed from astrophysics by not relying on the "minute details" of spectroscopy:

---

[96] Thomson 1911, p. vii.

[97] Clerke 1885, p. 183.

[98] Ibid. Clerke also referred in her book to "the part played in cosmical physics by chemical affinities" (p. 132), by which she meant the recognition that chemical combination and dissociation processes must play a role in the Sun's photoshere. She ascribed the recognition to the French astronomer Hervé Faye in 1865.

[99] Together with Peter Guthrie Tait, Stewart published in 1875 *The Unseen Universe*, a controversial book in which they attacked materialism and argued for a reconciliation of science and Christian religion. For other aspects of Stewart's engagement in the science-religion question, see Gooday 2004.

> "Cosmical physics" is a term broadly applied to the totality of those branches of science which treat of cosmical phenomena and their explanation by the laws of physics. It includes terrestrial magnetism, the tides, meteorology as related to cosmical causes, the aurora, meteoric phenomena, and the physical constitution of the heavenly bodies generally. It differs from astrophysics only in dealing principally with phenomena in their wider aspects, and as the products of physical causes, while astrophysics is more concerned with minute details of observation.[100]

The International Congress of Physics that convened in Paris in 1900 in connection with the World Exhibition that year included among its subjects *physique cosmique*.[101] Of the 92 invited papers published in the proceedings, nine belonged to the field of cosmical physics, the contributions ranging from gravimetric measurements over atmospheric electricity to polar light and the physical constitution of the Sun. The field was represented by, among others, Kristian Birkeland from Norway, Franz Exner from Austria, Roland Eötvös from Hungary and Adam Paulsen from Denmark. The absence of contributions to the congress in this area from British or American scientists did not mean that they were unconcerned with it. The following year J. J. Thomson gave a Royal Institution lecture on his view of the atom as made up of electrons. He reviewed sympathetically the cosmic effects of electrons such as expounded by Birkeland and Arrhenius, suggesting that the particles "may play an important part in cosmical as well as terrestrial physics."[102]

Another indication of the growth of interest in subjects of cosmical physics may be illustrated by the British Association for the Advancement of Science, which included a Section A on the mathematical and physical sciences. In 1898, the section included meteorology, but not astronomy, as a subsection. Many astronomers felt that their science was insufficiently represented, and as a result of their dissatisfaction a new subsection on astronomy appeared in the British Association meeting in 1900. Only two years later, the name of the subsection was changed to "Astronomy and Cosmical Physics." Arthur Schuster, a German-born physicist trained under Helmholtz and Maxwell, succeeded in 1888 Stewart as professor at

---

[100] Entry "Cosmic" (http://www.1911encyclopedia.org/Cosmic). The same entry appeared in the editions of 1922 and 1926.
[101] Guillaume and Poincaré 1900. See also Kragh 1999, p. 17. Staley 2008, pp. 166-204, deals in detail with the International Congress of Physics, but without paying attention to the section on cosmical physics.
[102] Thomson 1901, p. 416, and also in *Popular Science Monthly* 59 (1901), 323-336.

Owens College or what in 1903 was renamed the Victoria University of Manchester. He served as chairman of the new subsection at the Belfast 1902 meeting. "It was decided," he said, "to hand over to the already established subsection of Astronomy, other subjects, such as Meteorology, Terrestrial Magnetism, Seismology, and, in fact, anything that the majority of physicists are only too glad to ignore."[103]

Schuster used much of his address to warn against the obsession with observations in meteorology and cosmical physics generally. "Science is not a museum for the storage of disconnected facts and the amusement of the collecting enthusiast," he lectured, suggesting that only if this were kept in mind would cosmical physics become a proper science. The observational sciences needed to stay in close contact with their mathematical and experimental sisters. If so, "Cosmical Physics may remain an integral portion of Section A, and, though we acknowledge our weaknesses, we claim to have also something to teach." Schuster was as interested in cosmical physics as Stewart had been. In a letter to Rutherford, who succeeded him as professor in 1907, he wrote that he wished to write a book on cosmical physics, such as he had long planned.[104]

At the 1903 meeting in Southport, the name "cosmical physics" was for unknown reasons replaced by "meteorology." The address of the new chairman of the subsection, William Napier Shaw, was indeed on meteorology, but it also referred to cosmical physics, which Shaw took to be "the application of the methods and results of Mathematics and Physics to problems suggested by observations of the earth, the air, or the sky."[105]

The meeting of 1904 in Cambridge was a high point of cosmical physics within the British Association. According to William Lockyer, the subsection was characterized by "great vitality."[106] It was attended not only by many British scientists, including Schuster, H. Frank Newall and J. Henry Poynting, but also by distinguished foreigners such as Birkeland from Norway and Henry Norriss Russell from the United States. Whereas Birkeland, the leading expert in the physics of the aurora borealis, spoke on the relationship between sunspots and aurorae, the astrophysicist Russell gave an address on the masses and spectra of stars. As *The Observer* commented, there was no agreement over the nature and place of cosmical physics: "The Subsection of Astronomy and Cosmical Physics has kept its cumbrous title unchanged for another year, after high debate between those who suggest that Astronomy proper is not Cosmical Physics in the ordinary sense of the word Physics, but, if anything else than Astronomy, is Applied Geometry!"[107]

---

[103] Schuster 1902, p. 512. Without using the name, Schuster reviewed cosmical physics in a lecture given at the University of Calcutta in 1908 (Schuster 1911, pp. 118-157).
[104] Schuster to Rutherford, 7 Sep 1906, in Kargon 1977, p. 233. The book remained unwritten.
[105] Shaw 1903, p. 541. From 1900 to 1920, Shaw directed the Meteorological Office, which he turned into a more scientifically oriented institution.
[106] Lockyer 1904.

During the following decade cosmical physics remained as a subsection of Section A, sometimes alone and at other times conjoined with astronomy. At the 1911 meeting cosmical physics and meteorology each appeared with their own subsections, whereas astronomy had been transferred to a new subsection called "General Physics and Astronomy." In 1917 the British Association appointed a committee to advance and coordinate branches of geophysical research, with the meetings of the committee being handed over to the Royal Astronomical Society.[108] The subjects discussed in the committee meetings covered many of the subjects of cosmical physics, including terrestrial magnetism, the constitution of the atmosphere, aurora borealis, geodesy, seismology and the interior of the Earth.

Cosmical physics was in vogue in the first decade of the twentieth century, such as asserted by *The Times* in an article of 1904:

> "Cosmical Physics" is a department of science which is rapidly growing in favour. As a subsection of the British Association, for instance, it is scarcely three years old, owing its birth to the fusion of the existing subsection of astronomy with non-existing sections for meteorology, seismology, and magnetism. But once born it has justified its existence by growing rapidly and healthily, and has apparently fascinated the readers of papers and addresses in the main section.[109]

Radioactivity was among the startling discoveries of fin-de-siècle physics that contributed to the apparently healthy development of cosmical physics. If the energy of the stars were due to radioactive processes, as widely believed, didn't the phenomenon provide yet another link between the heavens and the Earth? According to the young British meteorologist George Simpson, the new rays had proved useful in explaining "a number of difficulties connected with cosmical physics, for example, the source of the sun's energy and comets' tails."[110] Moreover, he suggested that they might also explain the difference in electrical charge between the atmosphere and the surface of the Earth.

---

[107] Anon. 1904, p.360.
[108] Anon. 1917. Dreyer and Turner 1923, pp. 227-229.
[109] *The Times*, 27 December 1904, as quoted in *Observatory* 28 (1905), 119.

The fascinating new field of cosmical physics appealed to the public and also to many meteorologists. On the other hand, it was not taken very seriously by British physicists and astronomers, most of whom preferred to think of cosmical physics as just astrophysics. When Hugh Frank Newall, professor of astrophysics at Cambridge University, offered his view on cosmical physics in 1909 when retiring as President of the Royal Astronomical Society, he spoke of solar physics, comets and meteors.[111] He had no more to say about meteorology and geological phenomena than Clerke more than twenty years earlier. Perhaps the American geophysicist Louis Agricola Bauer was more correct in his evaluation than *The Times*. According to him, the establishment of a subsection on cosmical physics "did not apparently meet with the favor of the physicists themselves." As he saw it, "Our British colleagues want the cosmical physicists to stay with them and not flock off by themselves, and the present tendency seems, accordingly, to be at the British Association, not to form such a subsection."[112]

## 9. The United States

The idea of a combined astronomy and geophysics was not foreign to scientists in the new world. The American meteorologist and astronomer Cleveland Abbe, director of the Cincinnati Observatory and the father of weather forecasting in the United States, formulated in 1866 an ambitious plan of a "practical astronomy" embracing meteorology, geology, chemistry, physics and mathematics. He wanted an institution that covered research fields such as meteorology (as distinct from climatology), terrestrial magnetism, the tides, the temperature within the crust of the Earth, geodesy, earthquakes and more.[113] His vision of an astronomical geophysics came too early to include the field of astrospectroscopy and remained, at any rate, a vision.

---

[110] Simpson 1904. George Clarke Simpson (1878-1965) was in 1905 appointed lecturer of meteorology at Manchester University, the first position of its kind at a British university. For the role of radioactivity in astronomy and the earth sciences in the early twentieth century, see Kragh 2007.

[111] Newall 1909. The title of the address was due to the editors of *The Observatory*. Since its foundation in 1913 and until his retirement in 1928, Newall served as Director of the Cambridge Solar Physics Observatory. As Vice President of Section A, the British-American astronomer Ernest William Brown gave an address at the 1914 British Association meeting. It was reproduced in *Science* (vol. 40, 1914, pp. 389-401) under the title "Cosmical Physics." Yet the address was purely on the theory of the motions of the Moon, Brown's favourite subject, and not at all about astrophysics or cosmical physics.

[112] Bauer 1909, p. 566. Bauer, who had written his doctoral dissertation on terrestrial magnetism in Berlin in 1895, attended the 1909 British Association meeting in Winnipeg, where he gave a paper on geomagnetism.

Although astrophysics had become an American speciality already at the end of the nineteenth century, apparently there was little interest in its extension to a cosmical physics covering also terrestrial phenomena. The discovery in the late 1890s of the group of noble gases from helium to xenon was of interest to both astrophysicists and physical meteorologists. According to an American astrophysicist the presence of these gases in the atmosphere of the Sun had "an importance for cosmical physics that can hardly be overestimated."[114] He was referring to the stars, not to the atmosphere of the Earth. Realizing that meteorological research had to include not only the Earth, but also the Sun, Frank Bigelow, professor of meteorology in the Weather Bureau from 1891 to 1910, was more receptive to the ideas of cosmical physics. He dealt in several papers with the relation of solar magnetism, which he represented as a type of radiant energy, to geomagnetism and meteorology. The atmospheres of the two celestial bodies were bound together in what he called a "single cosmical thermal engine." Bigelow's argument for a "cosmical meteorology" was of the same kind that Stewart had suggested much earlier: "Solar physics and astrophysics are evidently only other names for meteorology, which embraces all atmospheric phenomena in its scope."[115]

In the St. Louis International Congress of Arts and Science held in September 1904, many of the addresses given to the sections of the physical sciences dealt with the new discoveries in electron physics and radioactivity and their consequences for the physical world picture.[116] High-profile European invitees such as Henri Poincaré, Ernest Rutherford, Wilhelm Ostwald and Ludwig Boltzmann helped making the congress a memorable event. Among the invited speakers was also Svante Arrhenius, who came with a fresh Nobel Prize to give an address to the section on "cosmical physics" placed under the earth sciences.[117] This section, which was originally named "meteorology," included two invited speakers, both Americans, in addition to Arrhenius: Abbott Rotch, Director of the Blue Hill Meteorological Laboratory near Boston, and Louis Bauer, Director of the Department of Terrestrial Magnetism at the Carnegie Institution.

---

[113] Reingold 1964. Cannon 1978, p. 77.
[114] Mitchell 1903, p. 227.
[115] Bigelow 1904, p. 30.
[116] Some of the key papers in these areas are reproduced in Sopka 1986. See also Davis 1904 for a general description of the congress.

Arrhenius' talk on "The Relation of Meteorology to other Sciences" aroused much attention and was followed by a large audience. "Not the least interesting feature of the address was the genial personality of the great Swedish philosopher himself," noted *The Times*. "Before the end of his address the room, which was well filled at the beginning, was closely packed with an eager audience."[118] *Popular Science Monthly* was no less impressed:

> The section of cosmical physics was … remarkable for the ideals of synthesis and the spirit of cooperation which pervaded it. In an address as bold as it was original Arrhenius proposed a theory of the possible connection between phenomena the most diverse and separated by exceedingly great distances, thus, *e.g.*, raising meteorology to the dignity of a cosmic science.[119]

Arrhenius offered a broad survey of meteorology, including such topics as Vilhelm Bjerknes' circulation theory for the atmosphere and his own theory of the "hothouse action" (later to be renamed the greenhouse effect) caused by atmospheric carbon dioxide. He also considered the physical conditions of the higher strata of the atmosphere, the temperature of the Sun's corona, the principal spectral line of the aurora borealis, and the chemical processes on the boundary between the atmosphere and the solid crust of the Earth. In these and other areas, "the sciences of meteorology, physics, likewise geology or its companion sciences, botany and zoology, work together."[120]

Although Arrhenius pointed out that solar physics and meteorology were in several ways closely connected, he did not consider astronomy and astrophysics in his presentation. He concluded "that meteorology not only stands in the closest

---

[117] The earth sciences at the St. Louis congress comprised eight sections: geophysics; geology; palaeontology; petrology and mineralogy; physiography; geography; oceanography; cosmical physics. The astronomical sciences comprised only two sections: astrometry and astrophysics.
[118] *The Times*, 27 December 1904.
[119] Davis 1904, p. 25.
[120] Arrhenius 1906, p. 737.

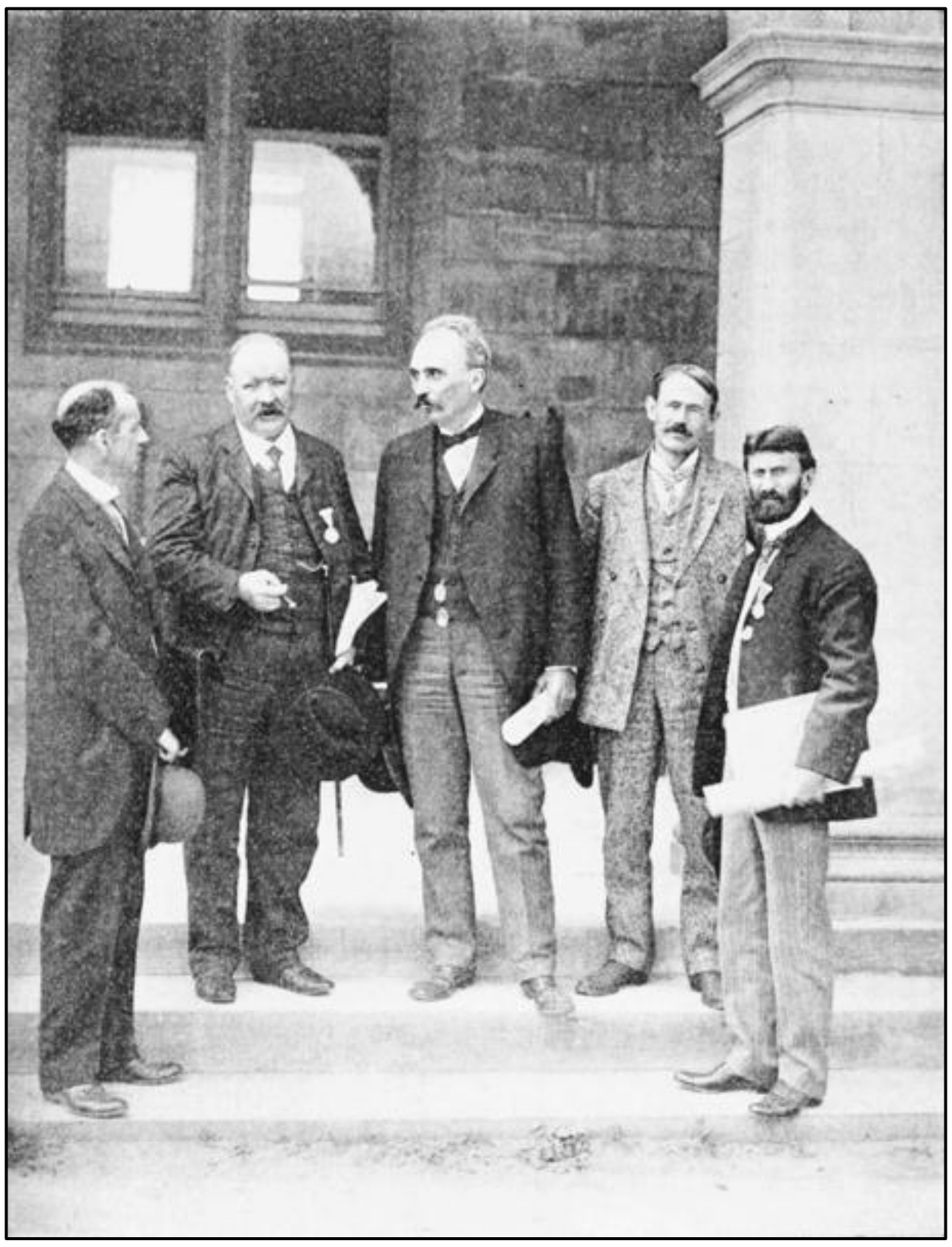

"Professors of the cosmical physics at the St. Louis Congress" as shown in *Popular Science Monthly* (Davis 1904). Abbott L. Rotch and Svante Arrhenius are to the left, Louis A. Bauer to the right.

connection with other branches of physical science as well with hydrodynamics, but that it is also connected intimately with questions of chemical, geological, and biological character." His address was indeed synthetic and broad-ranging, but what the audience listened to was closer to cosmical meteorology than to cosmical physics in its full meaning. After all, it was given to a section focusing on meteorology. As to the addresses on astronomy and astrophysics at the St. Louis congress, none of them referred to the earth sciences except in passing.

On the other hand, in his principal address to the section of geology the American geologist Charles Van Hise took a view that was no less broad and synthetic than Arrhenius'. Without using the term "cosmical physics" he proposed a reshaping of geology along the lines of cosmical physics and chemistry. Describinggeology as "one of the children of astronomy," he suggested that it was even closer related to physics and chemistry. Physics was the science of energy, chemistry the science of matter, and the two sciences were applicable all over the universe, including the Earth: "The most fundamental problem of geology [is] the reduction of the science to order under the principles of physics and chemistry. To a less extent geology is subject to the sciences of astronomy and biology."[121] Van Hise had earlier argued that the earth sciences should be seen as part of a larger and unified framework that included both the natural and the physical sciences. He referred in this regard to the new and progressive interdisciplinary fields of astrophysics and physical chemistry:

> For a long time astronomy and physics were pursued as independent sciences. The great discoveries of astro-physics have shown the advantages of their combination. Chemistry and physics for a long time were pursued as independent sciences. The rapid rise of physical chemistry has shown how wonderfully fruitful is the ground between the two.[122]

He envisaged yet another new interdisciplinary field – a "great, almost untouched territory" – between chemistry, physics and geology.

Louis Bauer argued in his address in St. Louis that astronomers and physicists should accord greater recognition to the study of terrestrial magnetism and terrestrial electricity, which he considered to be central to "the study of the physics of the universe."[123] As he made clear in another address a few years later, he was not happy with the development in the physical sciences, which had led sciences "classified under physics of the globe, meteorology, geodesy and navigation" to have a lower reputation than those of physics, mathematics and astronomy. Bauer sensed that the project of cosmical physics was falling apart, at least in his own country, being separated in superior and inferior parts:

> We note with pleasure that the American physicist is very prominently represented, indeed, in astronomy and astrophysics. May we not hope that he will soon realize that this planet on which we dwell, and which must form the basis of all our astronomical speculations, is also worthy of the highest and most unselfish devotion?[124]

---

[121] Van Hise 1904, p. 592. Van Hise did not consider meteorology or its relation to geology. On Van Hise and his role in American geochemistry, see Servos 1983.

[122] Quoted in Servos 1990, p. 234.

[123] Bauer 1906, p. 756.

[124] Bauer 1909, p. 569. Paper presented at the Baltimore 1908 meeting of the American Association for the Advancement of Science. It was also published in *Monthly Weather Review* 37 (1909), 27-29. Stephen Brush noted the inferior status that planetary science was assigned

The feeling that Bauer expressed, that meteorology and the other earth sciences was considered an underdog relative to physics and astronomy, was not new. In his report to the 1903 British Association meeting, Shaw gave voice to the same feeling, only more directly and with more bitterness. His words about the rivalry between scientific disciplines deserve to be quoted at some length:

> There was a time when Meteorology was a recognised member of the large physical family and shared the paternal affection of all professors of Physics; but when the poor nestling began to grow up and develop some individuality electricity developed simultaneously with the speed of a young cuckoo. The professors of Physics soon recognised that the nest was not large enough for both, and … that science was ejected as an Ishmael. Electrical engineering has an abundance of academic representatives; brewing has its professorship and its corps of students, but the specialised physics of the atmosphere has ceased to share the academic hospitability. So far as I know the British universities are unanimous in dissembling their love for Meteorology as a science, and if they do not actually kick it downstairs they are at least content that it has no encouragement to go up. In none is there a professorship, a lectureship, or even a scholarship, to help to form the nucleus of that corps of students which may be regarded as the primary condition of scientific development.[125]

The low regard with which many physicists considered meteorology, and the earth sciences generally, was not restricted to the Anglo-American scene. Meteorology in the fin de siècle era was not a science that leading physicists had much respect for. In the late 1920s Vilhelm Bjerknes recalled a conversation he had with the Berlin physicist Friedrich Kohlrausch some three decades earlier. Upon expressing his new interest in meteorological physics, Kohlrausch warned: "A physicist who goes into meteorology is lost."[126]

---

in the first half of the twentieth century, and also that in the earlier century the situation was quite different (Brush 1996, pp. 42-46). Bauer likewise pointed out that in the nineteenth century giants such as Helmholtz and Kelvin engaged whole-heartedly in the more "impure" fields of terrestrial and cosmical physics.

[125] Shaw 1903, p. 542. In an obituary of Shaw, E. E. Gold described how he and a few other physicists, including Schuster and George Darwin, strove "to weld the geophysical sciences into a coherent whole and prevent their relegation to a garret in the mansion of laboratory physics." *Quarterly Journal of the Royal Meteorological Society* 71 (1945), 192-194, on p. 192. On the emergence of British geophysics and its strong ties to astronomy (rather than astrophysics), see Kushner 1993.

[126] Quoted in Friedmann 1989, p. 48.

In the same year that the St. Louis congress took place, Satoru Tetsu Tamura of the Carnegie Institution's Department of Terrestrial Magnetism reflected on the future of meteorology and the need to explore the upper strata of the atmosphere, which he considered to be of ultimate importance "for the advancement of cosmical physics." Tamura felt that the status of meteorology as a science was low and that part of the reason was the absence of precision experiments and advanced mathematics: "Any student of mathematical physic [sic] who peeps into this adjoining field must feel that he is like a bat in the dark, flying at the gleams of light from a closely curtained window."[127] His remedy for improving the status of meteorology was to improve the mathematical training of young meteorologists, not to strengthen the connections between the discipline and neighbouring disciplines such as astrophysics and geophysics.

## 10. Cosmical physics after 1920

The project of cosmical physics barely survived the Great War and then only in a fragmented and weak state that included only a single university chair. After Exner had left the Innsbruck position in 1916, the chair remained vacant until 1927, when Arthur Wagner was appointed extraordinary professor in cosmical physics. Significantly, during Wagner's period as professor, the Institute of Cosmical Physics was officially renamed the Institute of Meteorology and Geophysics.[128]

No textbooks in cosmical physics were published in the interwar period and only a single institution seems to have been formally associated with cosmical physics. Victor Conrad, the former professor of cosmical physics at the University of Czernowitz, made an attempt to revive the field after he returned to Vienna to work for the Seismological Service of Austria. In 1926 he took over the editorship of *Gerlands Beiträge zur Geophysik* and a few years later he established a supplementary series called *Ergebnisse der kosmischen Physik*. The first volume, edited by Conrad and the Leipzig geophysicist Ludwig Weickmann, appeared in 1931. It contained lengthy review articles on issues such as the aurora, cosmic rays, atmospheric ozone, geodesy and the propagation of explosion waves in the Earth's atmosphere (sometimes known as "air seismology").[129] Among the contributors were highly recognized scientists including Carl Størmer, Walther Kohlhörster and Felix M. Exner.

[127] Tamura 1904, p. 464.
[128] Oberkofler 1971, p. 146.

A second volume of the *Ergebnisse* appeared in 1934, a third on atmospheric physics in 1938, and the fourth volume of 1939 was devoted to the physics of the hydrosphere and the lithosphere of the Earth. The book series thus dealt with central parts of cosmical physics, but largely restricted to areas of geophysics and atmospheric physics. Contrary to the older cosmical physics, as defined by the content of the textbooks by Arrhenius and Trabert, Conrad's series avoided astronomical, astrophysical and cosmological areas of research.

These aspects were not ignored in the 1920s and occasionally they appeared under the label of cosmical physics, although then with no concern for questions of geophysics. For example, when James Jeans in 1926 delivered a lecture on "Recent Developments of Cosmical Physics," it was strictly on problems of astrophysics, including such topics as the interior of stars, the sources of stellar energy, the temperature of stars and the still badly understood cosmic rays.[130] He had nothing to say about the Earth and its atmosphere. Again, when Lord Rayleigh (Robert John Strutt) three years later spoke on "Problems of Cosmical Physics" in his presidential address to Section A of the British Association at its meeting in Cape Town, South Africa, he essentially dealt with the spectroscopy of stars, nebulae and the aurora borealis.[131] In neither of the addresses were there any indications of the former, more comprehensive meaning of cosmical physics.

The few cases in which the term was used in the 1920s, it was associated with different meanings. For example, in a paper of 1923 the British-American physicist William Swann dealt at length with "Unsolved Problems of Cosmical Physics" without referring to astronomical issues.[132] The problems he included were atmospheric electricity and conduction, the maintenance of the Earth's charge and the origin of its magnetism, and gravitation. Neither Swann nor Jeans and Rayleigh mentioned the older tradition in cosmical physics with its aim of presenting a synthetic perspective on the earth sciences and their relations to solar and planetary physics.

---

[129] Conrad and Weickmann 1931. The volume received positive reviews in British journals of the earth sciences. See *The Geographical Journal* 79 (1932), 522-523 and *Quarterly Journal of the Royal Meteorological Society* 58 (1932), 313-314.

[130] Jeans 1926.

[131] Rayleigh 1929 and with same title in *Report of the British Association for the Advancement of Science 1929*, pp. 38-50. Still at the 1929 meeting the British Association included a subsection or "department"on cosmical physics.

[132] Swann 1923.

The only institution in the interwar period carrying the name "cosmical physics" was, to my knowledge, Norwegian. Based on a donation from the International Education Board, a branch of the Rockefeller Foundation, in 1928 an "Auroral Observatory" was established in Tromsø in northern Norway.[133] The aim of the observatory was to study the aurora borealis, atmospheric electricity, terrestrial magnetism and allied phenomena. Together with the Magnetic Bureau in Bergen, the Tromsø observatory was included in what was named the Norwegian Institute of Cosmical Physics (Norske Institutt for Kosmisk Fysikk). The institute lasted until 1972, when the University of Tromsø was established.

Interest in cosmical physics, including the name, reappeared in the German Democratic Republic in the 1960s, where the German Academy of Sciences in Berlin established a research area in *kosmische Physik* headed by the distinguished theoretical physicist Hans-Jörgen Treder, a specialist in the general theory of relativity. Following the 1989 reunification, in 2000 a German Commission for History of Geophysics and Cosmical Physics was founded under Treder's leadership. The commission publishes a journal edited by Wilfried Schröder, named *Beiträge zur Geschichte der Geophysik und kosmischen Physik*.[134]

## 11. Summary and discussion

It is difficult to pin down a precise meaning of the fin-de-siècle phenomenon known as cosmical physics, both when it comes to its nature and content. It was a kind of interdisciplinary research programme aimed at synthesizing and coordinating those parts of astronomy and the earth sciences that could be understood from the perspective of laboratory physics and, more generally, the laws of physics. While interdisciplinary, it was not itself a discipline. Perhaps it is better characterized as an "umbrella discipline" or a framework for studying the Earth by taking into regard its status as a planet within the solar system.[135]

---

[133] Friedmann 1995, pp. 30-33.
[134] http://verplant.org/history-geophysics/english.html
[135] See Whitley 1976, who speaks of an umbrella discipline as a loose disciplinary organization in which research is predominantly done at the specialty level without direct reference to, or influence from, the discipline.

During the first decade of the twentieth century cosmical physics obtained semi-official recognition as a scientific unit or group, such as indicated by its presence in abstract journals, encyclopaedia and the meetings of the British Association. It also figured in international contexts, including the 1900 world physics conference in Paris and the 1904 St. Louis congress. On the other hand, with the exception of Förster's *Mitteilungen*, mostly aimed at amateur astronomers in Germany and Austria, there was no scientific journal devoted to cosmical physics. Scientists working in the area saw no need for such a journal, as they could publish their research in existing journals of physics, meteorology, astronomy and terrestrial magnetism. Again with the exception of Förster's society for friends of astronomy and cosmical physics, there were also no professional societies with cosmical physics as their focus area.

The field could boast of three extensive textbooks – Müller's, Arrhenius' and Trabert's – but it is unclear how much they were actually used and in which contexts. Rather than forming a textbook tradition, they appeared as a series of independent works.[136] Regular courses in cosmical physics were limited to Austria, particularly at the University of Innsbruck, and it was also mainly in the Austro-Hungarian Empire that cosmical physics became institutionalized in the form of chairs and small institutes. However, in most cases the subjects dealt with at the Austrian institutions were terrestrial and atmospheric physics rather than cosmical physics in its broader meaning. Aspects of astronomy or solar physics played no significant role. Finally, there were no attempts to create a research school in cosmical physics, neither in Austria nor elsewhere. Indeed, there was no leader of the field, no individual scientist or group of scientists who felt committed to it and sought to promote it by means of, for example, a scientific journal.

It is evident in retrospect that one of the weak points in the very construction of cosmical physics was disagreement over its content, although the issue was never explicitly discussed. Scientists had widely different conceptions of what cosmical physics was or should be about, differences that typically reflected their backgrounds in either the astronomical sciences or the earth sciences. To Hann, Pernter and others in the Austrian tradition, the core fields of cosmical physics were meteorology and geomagnetism, whereas they tended to disregard astrophysics. Although Förster had a genuine interest in the impact of the Sun on the Earth's atmosphere, in his

136 On textbooks and discipline-building, see Gavroglu and Simões 2000.

conception of cosmical physics astronomy counted more heavily than meteorology. The same was the case with astronomers generally, to the extent they cared for cosmical physics at all – and most did not. Only very few scientists, notably among them Arrhenius in Sweden and Birkeland in Norway, were devoted to cosmical physics in its wide sense that took the word "cosmical" seriously. On the other hand, their excursions into cosmology and cosmogony attracted more public than scientific interest.

As to the physicists, not a few of them worked in areas that belonged to cosmical physics, such as atmospheric electricity and solar physics. However, during the same period that cosmical physics promised to develop into an important branch of science, fundamental physics was in a state of transition. Parts of the new physics, including cosmic rays and aspects of radioactivity, could be incorporated into the cosmical physics programme, but as a whole the new discoveries and theories constituted a centrifugal force that tended to make it less attractive and less necessary. The hot fields in physics lied elsewhere. As Schuster cynically remarked, the majority of physicists were "only too glad to ignore" the subfields making up cosmical physics.

Cosmical physics was a heterogeneous collection of fields, described by Trabert as "a loose juxtaposition of astrophysics, the physics of the solid and fluid parts of the Earth, and the physics of the atmosphere." If that were all there was to it, there was no obvious reason why the fields should be pursued under a common framework. Moreover, the research programme tended to be internally unstable, in so far as it consisted of disciplines of a very high scientific status, such as physics and astronomy, and others of a low status, such as meteorology. This problem, highlighted by several meteorologists, contributed to the dissolution of the cosmical physics project.

When the International Union for Geodesy and Geophysics was founded in 1919 it comprised six earth sciences, namely, geodesy, meteorology, terrestrial magnetism and electricity, seismology, physical oceanography, and the study of volcanoes. Cosmical physics was not included and neither did it appear in other of the unions of the newly established International Research Council. In a study of the disciplinary formation of geophysics, Gregory Good has advocated that disciplines should be regarded as ever-changing frameworks within which a diversity of scientific work is organized and conducted.[137] In the case of cosmical physics there was a framework, but not much more. Scientific research dealing with the fields encompassed by the proposed project did not take place within the framework but independent of it. To a large extent, cosmical physics remained a label rather than a real organizational unit for doing science. And, as Good points out, use of a term indicates only a desire to designate a new field, not success in doing so.

Geochemistry is another interdisciplinary branch of science to which cosmical physics may be compared. Although with roots in the nineteenth century, it was only after World War II that geochemistry became a scientific discipline with its own journals, institutions, chairs, textbooks and professional societies. An important factor in the formation of the discipline was the founding in 1950 of the international journal *Geochimica et Cosmochimica Acta*, signifying the recognition of cosmochemistry as a sister discipline of geochemistry. "The chemistry of the Earth and of the Cosmos has become a branch of science independent enough to have a journal of its own," the editors argued, and the success of the journal proved them right.[138] The aims of cosmochemistry resembled in some respects those of the earlier cosmical physics, but whereas the first field succeeded as a scientific discipline, the latter did not.

One may argue that fin-de-siècle cosmical physics was doomed to fail, not because the project of integrating geology and astronomy was inherently unrealistic but because it was premature. In the decades after 1920, many of the branches encompassed in the vision of a cosmical physics developed rapidly and largely independently, nourished by new discoveries that were outside the vision of the older framework. Cosmic rays physics, still at its infancy in 1920, turned into an exciting research area that heralded the coming of elementary particle physics. While astrophysics before World War I had still been largely empirical and limited to spectroscopy, in the 1930s it developed into a mature physical science with strong ties to the new nuclear physics. Similar if less dramatic progress occurred in the fields of geophysics.

It was only after the individual sciences related to astro- and geophysics had progressed sufficiently that a new and even more comprehensive kind of cosmical physics became a productive framework with results that were more than the sum of knowledge of the individual sciences making up the framework. This level was reached with the emergence of space science in the 1960s. Over the last half century the links between astronomy and geophysics have multiplied and intensified, to the advantage of both of the sciences. Although the name "cosmical physics" is no longer used, the dream of it has become a reality. And, contrary to the old days, it is an interdisciplinary project in which astronomers and physicists take as much pride and interest as meteorologists and geophysicists.

---

137 Good 2000. See also G. A. Good, "Disciplinary history," pp. 171-172 in Good 1998, and Schröder 1982.

138 Quoted in Kragh 2001, p. 181. Solar physics emerged as a discipline or subdiscipline in the 1960s. For the role of a new scientific journal (*Solar Physics: A Journal for Solar Research and the Study of Solar Terrestrial Physics*) in this process, see Hufbauer 1989.